\documentclass[a4paper,11pt]{article}
\usepackage{aaskaiid}
\usepackage{aas_macros}  
\usepackage{orcidlink}
\title{Evolved massive stars and their impact on their environment }
\ShortTitle{Evolved Massive Stars and Environment}

\author[1]{C.S.~Buemi\orcidlink{0000-0002-7288-4613}}
\ShortName{C.~Buemi et al.} 
\author[1]{G.~Umana\orcidlink{0000-0002-6972-8388}}
\author[1]{C.~Trigilio\orcidlink{0000-0002-1216-7831}}  
\author[1]{C.~Bordiu\orcidlink{0000-0002-7703-0692}}
\author[1]{F.~Bufano\orcidlink{0000-0002-3429-2481}}
\author[2]{J.~van~den~Eijnden \orcidlink{0000-0002-5686-0611}}
\author[3]{S.~Orlando \orcidlink{0000-0003-2836-540X}}
\author[1]{P.~Leto\orcidlink{0000-0003-4864-2806}}
\author[3]{F.~Bocchino  \orcidlink{0000-0002-2321-5616}}
\author[4,3]{M.~Miceli \orcidlink{0000-0003-0876-8391}}
\author[5]{M.~P{\'e}rez-Torres\orcidlink{0000-0001-5654-0266}}
\author[5]{A.~Alberdi\orcidlink{0000-0002-9371-1033}}
\author[1]{F.~Cavallaro\orcidlink{0000-0003-1856-6806}}  
\author[6]{L.~Cerrigone\orcidlink{0000-0002-5537-7134}} 
\author[1]{A.~Ingallinera\orcidlink{0000-0002-3137-473X}}
\author[1]{S.~Loru\orcidlink{0000-0001-5126-1719}}  
\author[1]{S.~Riggi\orcidlink{0000-0001-6368-8330}}  
\author[1]{A.C.~Ruggeri\orcidlink{0000-0002-1556-2474}}   

\affiliation[1]{INAF-Osservatorio Astrofisico di Catania, via S. Sofia 78, 95123, Catania, Italy}
\emailAdd{carla.buemi@inaf.it}
\affiliation[2]{Anton Pannekoek Institute for Astronomy, University of Amsterdam, Science Park 904, 1098 XH, Amsterdam, The Netherlands}
\affiliation[3]{INAF - Osservatorio Astronomico di Palermo, Piazza del Parlamento 1, 90134, Palermo, Italy}
\affiliation[4]{Dipartimento di Fisica e Chimica E. Segrè, Università degli Studi di Palermo,Via Archirafi, 36, 90123, Palermo, Italy}
\affiliation[5]{IAA-CSIC, Instituto de Astrofísica de Andalucía, Glorieta de la Astronomía s/n, 18008 Granada, Spain}
\affiliation[6]{Joint ALMA Observatory, Alonso de C{\'o}rdova 3107, Vitacura, Santiago, Chile}

\abstract{The comprehension of the final stages of massive star evolution and their path toward the eventual supernova explosion necessarily involves the study of their stellar winds and their circumstellar environment (CSE) in the transitional phases.  During these phases, mass loss (via stellar winds and eruptive events) profoundly shapes the surrounding medium.
The study of the pre-supernova progenitors, from Red Supergiants, passing through the Luminous Blue Variable stage to Wolf-Rayet stars, is of key importance because, focusing on their nebulae, they directly prove the mass-loss activity of the star that, through wind and eruptive events, shapes  the environment in which the supernova will explode. Such environment, interacting with the ejecta, will heavily affect  supernovae spectrophotometric signatures.
 
The Square Kilometre Array, with its extraordinary capabilities to combine high spatial resolution, sensitivity and wide frequency coverage will address the most critical observational issues that currently prevent detailed characterisation of CSE and, thus limit our ability to constrain its connections to supernova and remnants properties.
}

\begin{document}
\include{journal-names}

\maketitle

\section{Introduction}
 Massive stars lose a significant fraction of their mass during the complex phases of post-main sequence evolution, injecting energy, momentum, and matter into the interstellar medium. The mass-loss history determines the evolutionary path of such stars, affecting their final fate \citep{Smith_2014, Vink_22}. Mass can be ejected via steady stellar winds, episodic outbursts, and/or explosive events, and the way in which it occurs shapes both the immediate stellar surroundings and the larger interstellar medium. On even larger scales, in fact, the material ejected by massive stars enriches the interstellar medium (ISM) with heavy elements and kinetic energy. This process plays a crucial role in galactic evolution, influencing subsequent star formation and the chemical composition of galaxies \citep{Chisholm_18, Eldrige_22}. Despite this, the post-main-sequence evolution of massive stars remains among the least well understood stages of stellar life, and the mechanisms driving the observed instabilities are still unclear.
 
 In this context, the CSE  is not only a secondary effect, but is the only direct observational way to quantify the mass-loss phenomena that theoretical models struggle to reproduce with accuracy \citep{ Renzo_17,Dessart_24,Agrawal_22}, and to constrain the physical parameters (mass, metallicity, binarity) that determine the dynamics, luminosity and final classification of stellar transients \citep{Langer_12, Smartt_15}.
It constitutes an archive of the stellar mass-loss history, 
 and therefore its characterisation is fundamental to trace the instabilities and mechanisms that drive mass ejection throughout stellar evolution 
 \citep{Smith_2014,Chandra_18,leung_21}.
 
Progress in understanding must tackle both the inability of theoretical models to reproduce the complex transition from the main sequence toward the core-collapse supernova (SN) and SN remnant formation, hindered by the lack of an accurate mass loss characterisation \citep{Vink_22}, and the paucity of robustly classified objects in the most complex and unstable phases of late evolution.

However, radio observations have proven to be fundamental in understanding many aspects of
the evolution of massive stars, from their stellar winds to interstellar medium interactions
and the environment in which SN explosions occur \citep[e.g.][]{Umana_g26_12, Agliozzo_17,Usta2021}. 
The main advantage of observing continuous radio emission is that it does not suffer from strong absorption by the dense circumstellar nebulae,
allowing observations of  both ionised stellar winds and circumstellar structures. 

The potential, together with the greatly improved capabilities of next-generation radio facilities, is already becoming evident through observations with SKA pathfinder and precursor instruments. In these last few years, the Australian Square Kilometre Array Pathfinder (ASKAP) and MeerKAT have provided several new insights, revealing circumstellar nebulae around evolved massive stars, detecting radio emission from infrared bubbles or identifying new populations of compact radio shells (Umana et al. in prep.; 
\citealp{Bordiu_Kyklos_24, Bordiu_25}), and discovering new stellar radio bow shocks  \citep{Van_den_Eijnden_vela_22,Van_den_Eijnden_22}. These results represent a compelling preview of the scientific breakthroughs that the full SKA will deliver. 

This chapter focuses on SKA-Mid, whose frequency coverage optimally probes thermal emission from stellar winds and circumstellar nebulae; the role of SKA-Low in discriminating between thermal and non-thermal emission processes is also discussed.
We initially introduce the post-main sequence evolutionary stages on which the chapter focuses (Sect.~\ref{stages}),  describing how the radio continuum diagnostics, particularly spectral
index analysis and morphological studies, allow us to study both small-scale stellar processes and
large-scale nebular structures, as well as their imprints on supernova ejecta and remnants (Sect. ~\ref{radio_ins}). Section \ref{sec:snr} discusses how the CSM shaped by massive star mass loss affects supernova explosions and remnant properties. Section~\ref{limit_and_potential} examines the limitations of current radio facilities, showing through synthetic simulations how the combination of sensitivity, angular resolution, and wide frequency coverage offered by SKA will address them, providing observational support needed by theoretical models to link progenitor properties to their ultimate fate. Finally, we conclude (Sect.~\ref{synergies}) by discussing the importance of synergies of SKA radio observations with complementary multi-wavelength observations.

\section{Mass Loss in Advanced Evolutionary Phases of Massive Stars}
\label{sec:mloss}
\label{stages}
\subsection{Luminous Blue Variable: a Strongly Unstable Transitional Phase}
\label{sec:LBV}
Luminous Blue Variables (LBVs) are the post-main-sequence descendants of the most massive stars evolving toward the Wolf-Rayet stage \citep{Weis_20}; it has been proposed that in some cases they may directly precede gravitational collapse and directly experience the subsequent supernova explosion \citep{Kotak_06, Groh_13}.

This evolutionary phase is characterised by phenomena not yet well understood of radiative and dynamic instability that lead to extreme mass loss, both through steady stellar winds and/or intense and short-lived events, which define the stellar quiescent and eruptive phases. During the quiescent phase, mass loss occurs via dense and continuous wind, with a rate of $~
\dot{M} \sim 10^{-6} - 10^{-4} \, M_{\odot} \, \text{yr}^{-1}
$
and terminal speeds ranging from a few hundred to a thousand km/s, typically lower than those observed in Wolf-Rayet stars. 
The mass loss increases dramatically during the eruptive phase, when the star can eject up to \(10\, M_{\odot}\) in a few years, as observed in the prototypical case of $\eta$ Carinae during the 19th century eruption \citep{Morris_17}. Such giant eruptions have rarely been observed, but their effects are imprinted into the morphology and physical conditions of the circumstellar nebulae that are often observed around LBVs. These nebulae serve as direct witnesses of tens to thousands of years of mass-loss history, both before and during the LBV phase. \cite{Smith_17} and \cite{Weis_20} offer exhaustive reviews of the observational and evolutionary characteristics of such a class of objects.

Great uncertainty about the duration of this phase and the total amount of mass lost during the LBV phase prevents us from correctly reproducing the transition between O and Wolf-Rayet stars. Therefore, a full characterisation of the mass-loss properties during the LBV phase appears crucial to understanding their role in massive star evolution. 

The study of the central object together with the associated nebulae is a valuable tool to derive the total amount of gas (ionised, neutral, and molecular if it exists) and dust. Another important aspect of studying circumstellar envelopes is to determine the mass-loss history of the central star and, in particular, how the mass-loss behavior (multiple events, bursts) is related to the physical parameters of the central object \citep{Umana_iras_05, Umana_g26_12, Buemi_iras_10, Buemi_hr_17}.

High-resolution radio observations allow us to map such ejected matter, revealing complex structures such as bipolar shells, lobes, or equatorial rings, which could reflect the impact of hydrodynamic interactions and, in some cases, the presence of stellar companions \citep{Buemi_iras_10,Buemi_hr_17, Umana_g26_12}. The spectral characteristics of the radio continuum also make it possible to discriminate between phases of free expansion and ionised shock conditions, and thus provide constraints on expansion times and electron densities (Sect.~\ref{sec_cse}).

The study of the LBV population is in fast evolution thanks to the improvements in the observational capabilities, although it suffers from the paucity of confirmed members due, for example, to the heterogeneity of their brightness variations and to the different timescales on which they occur. 
Moreover, the current studies of their circumstellar nebulae are still mostly limited to our galaxy or Magellanic Clouds, due to instrumental resolution and sensitivity limitations. The ASSESS project \citep{Bonannos_iaus24} has successfully identified new B[e] supergiants and LBV candidates in nearby galaxies, particularly noting their presence in low-metallicity environments ($\sim$0.14 Z$_{\odot}$, \citealp{Maravelias_23, Bonannos_24A&A}), while recent millimetre studies are revealing the importance of the molecular gas component in the mass-loss budget and expanding the circumstellar molecular inventory \citep{Bordiu_19, Bordiu_20, Bordiu_22, Rizzo_23}.

\subsection{Wolf-Rayets: the Last Stages Before the Supernova}
\label{sec:wr}
Wolf-Rayet (WR) stars represent the final pre-supernova phase in the evolution of massive stars ($\gtrsim$20~M$_{\odot}$, \citealp{Massey_00,Crowther_07}), when the hydrogen-rich outer envelope has been completely removed by intense episodes of mass loss, often passing through LBV or Red Supergiant  phases. The remaining compact core, heavily enriched with nuclear products (He, C, N, O), is surrounded by an extremely 
dense and fast wind. The typical mass-loss rates of $\dot{M} \sim 10^{-5} - 10^{-4} \, M_{\odot} \, \text{yr}^{-1}$ and terminal velocities reaching 3000-5000 km/s \citep{Crowther_07}, make these stars among the most extreme wind sources known. These winds are often highly clumped rather than smooth and homogeneous \citep{Moffat_88, Lepine_2000, Lepine_08,Chene_20}, with important implications for the interpretation of radio observations, as discussed in Sect.~\ref{swind}.  In addition, in WR+OB binary systems, the interaction between the two stellar winds creates colliding-wind regions (CWRs) where the particles can be accelerated and produce non-thermal radio emission along with the thermal emission from individual winds \citep{Eichler_Usov1993}. 

Many WRs are surrounded by shells, bipolar structures, or ring-like nebulae, detected mainly through their dust emission in the mid-infrared \citep{Toala2011, Toala2015}. They likely result from the dynamic interaction between the current fast wind and the material ejected in previous evolutionary phases \citep{Garcia-Segura_96,zavala_22}. Such an interaction could generate shock fronts and compressed regions with enhanced density, temperature, and ionisation. Radio continuum observations can detect the ionised gas component of these nebulae when sufficiently dense and bright \citep{Cappa2004, Cohen_05, Burgemeister_13}.

Within the WR population, \cite{Smith_Conti} identified nitrogen-rich WR stars with residual hydrogen (WNh) as a particular class, characterised by ongoing core hydrogen burning.  Their  masses (>~50-60~M$_{\odot}$) and 
luminosity (\(\log L/L_{\odot} \gtrsim 6\)) are usually higher than those of the H-poor WRs, and their intense winds reach mass loss of  \(\sim 10^{-5} \, M_{\odot} \, \text{yr}^{-1}\). The evolutionary link between WNh stars and LBVs is still not well defined, with scenarios including a direct transition between the two phases, as well as the possibility of a reversible or cyclical evolutionary path \citep{Smith_96, Smith_Conti, Vink_2012}.
At radio wavelengths, WNh associated nebulae often show symmetric or multi-lobed structures, partly influenced by previous winds and partly sculpted by the radiative and mechanical feedback of the current high-energy wind. The thermal radio emission of these nebulae, combined with infrared and optical spectroscopic studies, allows reconstructing the recent history of mass loss and the transition from the initial OB stage to the terminal WR phase \citep{Burgemeister_13, Buemi_mgps_inprep}.

\subsection{Red Supergiants and Runaway Star Bow Shock}
\label{sec:bow_s}
Although less extreme, Red Supergiants (RSGs) represent an interesting phase of the evolution of massive stars with initial masses between about 8 to 40 M$_{\odot}$, and are considered to be the main progenitors of Type II supernovae \citep{Smartt_15}. Their mass loss is, in fact, associated with a slower wind than that of the WRs and LBVs, but it is still crucial in determining the structure of the pre-SN environment \citep{VanDyk_25}. However, discrepancies between the observed RSG progenitor masses and theoretical predictions raise questions about the evolutionary path followed by the most massive RSGs \citep{Smartt_09,Smartt_15}.

With their low electron density and temperature, RSG winds are only weakly ionised, making the radio detection of their thermal emission particularly challenging \citep{Harper2001}. As a result, direct radio observations of RSG winds have mostly been limited to nearby, high mass-loss rate stars like Betelgeuse, Antares, and VY Canis Majoris \citep{Gorman20, Lipscy_05}.

Some massive stars in advanced evolutionary phases, including some RSGs, become runaway stars that create detectable bow shock structures as they move supersonically through the ISM \citep{Blaauw_61}. The interaction of their winds with the surrounding medium often leads to large-scale structures such as detached shells, termination shocks, and bow shocks, particularly when the star is moving supersonically through the ISM. Bow shocks driven by runaway massive stars are mostly discovered in the IR band, where thermal dust emission renders them detectable in survey data and, in rare cases, also in optical (line) emission. Large-scale IR surveys have led to the compilation of extensive catalogs, such as the Extended Bow Shock Survey \citep[E-BOSS;][]{2012A&A...538A.108P, 2015A&A...578A..45P} and the Milky Way Project \citep{2016ApJS..227...18K}, which together contain hundreds of candidate bow shocks. These structures trace the cumulative effects of wind-ISM interaction and could provide crucial information on mass-loss history, stellar motion, and local ISM density. 

Radio emission, both thermal and non-thermal, from bow shocks was rarely detected before the advent of SKA pathfinder telescopes: only the bow shock driven by BD +43$^{\rm o}$3654 was detected with the Jansky Very Large Array (JVLA) \citep{2010A&A...517L..10B}, while since then a second JVLA-detected bow shock candidate was reported in NGC 7635 by \citet{2022A&A...663A..80M}. MeerKAT and ASKAP, through both pointed observations and survey data (e.g., ThunderKAT Large Survey Project, the Rapid ASKAP Continuum Survey; the Evolutionary Map of the Universe; \citealt{Fender_16,McConnell_20,Norris_11}) have proven highly efficient in detecting radio emission from  bow shock, reporting (candidate) radio bow shocks, and  bringing the total number to $\sim 10$ systems and strong candidates \citep{Van_den_Eijnden_vela_22, Van_den_Eijnden_22,  2024MNRAS.532.2920V}. Regarding the future opportunities enabled by the SKA, it is key to mention the recent breakthrough results revealed by Gaia DR3, showing that a large fraction ($\sim 25$\%) of O-type stars escape their birth cluster at supersonic velocities \citep{2024A&A...681A..21S,2024Natur.634..809S}. These results suggest that a large number of new runaways will be identified in the upcoming final Gaia data releases \citep[see also e.g.,][]{2023A&A...679A.109C}. Combined with these new runaway populations, deep SKA observations will be able to probe the presence and properties of bow shocks without the bias of prior IR shock detection. A quantitative discussion of the detectability of radio bow shocks and of the diagnostic power
of broadband SKA radio observations in disentangling their emission mechanisms is presented in Sect.~\ref{ska:bow_s}.

\subsection{Massive stellar clusters}
Massive stellar clusters serve as ideal laboratories for studying the properties and life cycle of massive stars, as they host stellar populations that span a wide range of evolutionary stages, all within the same local environment and age \citep{Portegie_10}. These environments provide the opportunity to identify and characterise the different evolutionary phases experienced by massive stars, as described in the previous sections. The coeval nature eliminates many of the uncertainties associated with distance, metallicity, and formation conditions that affect field star studies, allowing direct comparisons between different evolutionary stages and masses. The high stellar density within these clusters also creates unique physical conditions in which stellar winds interact not only with the ambient medium but also with each other, producing complex hydrodynamic structures and collective feedback mechanisms that cannot be studied in single star studies  \citep{Canto_00}.
Radio emission has been observed from Westerlund~1, the most massive and nearest stellar cluster to us \citep{Dougherty2010AA...511A..58D}. This emission is typically attributed to stellar winds from early-type and WR stars, as well as to circumstellar ejecta associated with massive stars (thermal), and wind-collision zones in interacting binaries (non-thermal). Recent MeerKAT observations have provided an unprecedented view of the cluster in the radio, revealing details and recovering extended low-brightness structures, never seen before, that appear to be moving away from the cluster’s center (Umana et al., in preparation). These features could be related to the feedback from the cluster. Large-scale feedback is also supported by recent $\gamma$-ray emission \citep{Aharonian_22}, which underscores the role of massive stellar clusters in the acceleration of highly energetic particles. Massive star clusters located at the Galactic Center (the Nuclear Star Cluster, the Arches Cluster, and the Quintuplet Cluster) have been studied using high-resolution radio observations \citep{Lang_05, Gallego_22}. These observations reveal evidence of free-free thermal radiation, which likely originates from stellar winds and non-thermal emission from colliding-wind binaries (CWBs), as indicated by their consistent flat-to-negative spectral index \citep{Cano-Gonzalez_24, Cano-Gonzalez_25}.
The high sensitivity, angular resolution, and dense inner-core array layout of SKA-Mid in Band 5 will allow for studies similar to those conducted on Westerlund~1, but across a variety of clusters with different masses and ages. This will enable the investigation of complex interactions between stellar winds, diffuse emission within and around the clusters, the presence of global cluster winds, and the distribution of material surrounding these clusters. In addition, it will explore their potential connections to $\gamma$-ray emission. Studies of massive star clusters located near the Galactic Center will be particularly valuable. Their unique location offers a rare opportunity to study how these phenomena are influenced by extreme environmental conditions.

\section{Radio Insights into Evolved Stellar Environments} 
\label{radio_ins}
\subsection{Stellar winds}
\label{swind}
The radio spectral index ($\alpha$, where flux density S$_\nu\propto\nu^{\alpha}$) is one of the fundamental parameters for the study of stellar winds in evolved massive stars. It provides direct 
information on wind structure, clumping, internal and external shocks, and the interaction between wind and circumstellar nebula. An analytical description of free-free emission from ionised expanding stellar winds has been provided by \citet{Wright_Barlow1975} and \citet{Panagia_Felli1975}. For a spherically symmetric isothermal wind with constant velocity $v_{\infty}$ and mass-loss rate $\dot{M}$, the radio flux density at frequency $\nu$ from a star at distance $d$ is given by the following relation \citep{Scuderi1998}:
\begin{equation}
    S_{\nu} = 7.26 \left( \frac{\dot{M}}{10^{-6} M_{\odot} \text{yr}^{-1}} \right)^{4/3} \left(\frac{1.3~v_{\infty}}{100 \text{ km s}^{-1}} \right)^{-4/3} \left( \frac{\nu}{10 \text{ GHz}} \right)^{0.6} \left( \frac{d}{\text{kpc}} \right)^{-2} \left( \frac{T_{\text e}}{10^4 \text{ K}} \right)^{0.1} ~\text{mJy}
\end{equation}
where
$T_{\text e}$ is the wind temperature. 
This formulation allows a robust estimation of mass-loss rates from radio observations in the absence of  non-thermal emission or strongly anisotropic wind structures. 
Nevertheless, LBVs and WRs have highly structured winds, with strong deviations from sphericity and homogeneous distribution \citep{Davies_05, Vink_22}. Instead of being uniform, the winds 
are clumped, consisting of high and low density regions, and the way in which this affects the flux levels and the emerging spectral 
energy distributions (SEDs) must be taken into account in the theoretical treatment. As discussed in the literature \citep{Puls_08, Ignace2016, Daley2016, Huang2023}, these effects depend on several 
factors, such as the frequency of observation, the radial distribution of the clump (density, size, spacing), the temperature of the wind, and whether the wind is optically thick or thin at the given frequency. In particular, for a given mass loss rate, the presence of small, optically thin clumps generally leads to emitting flux higher than expected from a smooth wind, especially in the optically thick regime where the structured 
wind results in a pseudo-photosphere at large radii \citep{Ignace2016, Daley2016}, due to the enhancement of local emissivity \citep{Flores_21}, with a scale factor that is a function of the volume filled by clumps \citep{Abbott_81, Ignace2016}. It can also affect the spectral index, as a non homogeneous radial distribution of clumps or changing of optically thin/thick regime has different effects at different frequencies.
 
Other factors could influence the radio flux and spectral index, such as a partial ionisation degree of the wind \citep{Drew_89}. The free-free opacity is reduced 
when the wind is not fully ionised since the opacity scales with the square of the electron density \citep{Leitherer_91}. The less ionised the gas is, the closer the 
effective emission region is to the star. In dense environments steeper than the canonical $r^{-2}$, the optical depth dependence on the frequency 
could become even stronger and result in a steeper spectral index (i.e. $\alpha \gtrsim 0.6$). As a result, the spectral slope itself may vary with frequency, becoming steeper at higher frequencies, which probe deeper and denser regions of the stellar wind. 
 Such dense environments with a different radial density profile may also be caused by a wind that has not yet reached its terminal velocity at the radio photosphere, as evidenced in radio/mm spectra of OB supergiants \citep{2025MNRAS.543..862V}.

All this is of great importance in the study of the winds of evolved massive stars, where the shocks originated by their unstable dynamics and 
strong speed gradients could impact the homogeneity and ionisation property of the wind, and thus the observed spectral characteristics. 

In WR+OB binary systems, the interaction between stellar winds introduces additional complexity. In the colliding-wind region (CWR), gas is compressed and heated, creating conditions for particle acceleration, as recently reviewed by \cite{White_Tuthill_26}. Here, non-thermal radio emission components may originate, with synchrotron radiation emitted by spiraling relativistic electrons in shock-intensified magnetic fields, resulting in a flat or decreasing radio spectrum ($\alpha \leq$ 0), intrinsic polarization and orbital variability. The orientation of the wind-wind collision region changes according to the system's orbital motion. The total intensity varies between the observing epochs, depending on the orbital separation. The synchrotron spectrum evolves through the observed orbital phases, exhibiting optically thin and optically thick emission depending on the relevant absorption and cooling mechanisms. At the highest angular resolution, the shape and position of the wind-wind collision bow shock relative to the stars reveal the momentum of the different wind components \citep{Dougherty_2005,Sanchez-Bermudez_2019}. The identification of the non-thermal component is of fundamental importance to properly derive the physical parameters of the WR wind: the synchrotron emission can, in fact, overlap and contaminate the thermal one, leading to an overestimation of \(\dot{M}\) if not properly modeled. In addition, the presence of non-thermal radiation provides a unique opportunity to investigate particle acceleration mechanisms in stellar environments, similar (on a smaller scale) to those found in supernova remnants and AGN jets. Evidence of free-free absorption modulated by the orbital phase also provides powerful three-dimensional (3D) diagnostics on wind geometry and system inclination, especially when integrated with optical, X-ray, and infrared data.

\subsection{Circumstellar Environment: Morphology and Spectral Index Analysis}
\label{sec_cse}
Nebula studies help to reconstruct the evolutionary stages of massive stars and the physical processes driving mass loss. The mid-IR is the ideal window to explore the dust component of the stellar ejecta, while the observations at radio frequencies allow to reveal their ionised gas component \citep[e.g.][]{Umana_g26_12, Buemi_hr_17}. The circumstellar nebulae created by the material ejected throughout the evolutionary stages around the massive stars record in their structure and physical conditions a wealth of information about the star's mass-loss history and its interaction with the environment, as comprehensively discussed by \citet{Vink_22} and \citet{Smith_2014}. The velocity structure of the ejected material, the characteristics of the surrounding interstellar medium, stellar rotation, magnetic fields, and the existence of binary companions are some of the factors that influence the resulting architecture \citep{Langer_12,van_Marle_15}. 

The complexity and turbulence of the post-main sequence path of massive stars, closely linked to the transition between different phases of stellar wind, is thus reflected in a multiplicity of morphologies observed in their surroundings \citep[e.g.][]{Weis_20,Toala2015}.
Nearly spherical nebulae represent the simplest morphological class, generally associated with a relatively isotropic mass loss, and are often observed around some evolved massive stars during phases of steady mass loss. Their radio emission can be modeled using relatively simple geometric assumptions, as well as the shell structures, with intensified emission along the edges of the nebula that could indicate stellar wind interaction with previously ejected material or with the surrounding interstellar medium \citep{Toala2011}. The interaction between stellar winds at different velocities and ages creates regions of compression and shock which manifest themselves as localized increases in radio thermal emission (Sect.~\ref{sec:wr}). These interactions give rise to hydrodynamic instabilities which produce filamentous structures, intensified emission nodes, and complex morphologies observable in high resolution radio maps \citep{Garcia-Segura_96, Buemi_inprep}.

In LBVs, the formation of ring structures often reflects the episodic and violent nature of their mass loss events: during relative quiescence phases, the less intense stellar wind interacts with the previously ejected material during outburst events, creating an annular compression zone where radio emission intensifies considerably. 
However, as introduced in Sect.~\ref{sec:LBV}, more complex morphological structures are often observed, such as asymmetric or bipolar nebulae, whose 
formation is typically attributed to intensified mass loss along the stellar equator combined with lower mass-loss rates in polar directions, potentially driven by stellar rotation, magnetic fields, or binary interactions.

The interaction between circumstellar nebulae and the surrounding medium creates additional complexity in their structure and evolution, due to 
the formation of shock fronts and the deceleration of the expanding material \citep[e.g.][]{Agliozzo_17,Buemi_hr_17}. 
Similarly, the bow shock structures formed by runaway massive stars, where stellar winds interact with the ambient medium to create extended shock fronts, can exhibit both thermal and non-thermal radio emission. The complementary roles of SKA-Low and SKA-Mid in probing the extended and compact shock structures are addressed in Sect.~\ref{ska:bow_s}. The analysis of the intensity and spectral properties of radio continuum emission allows to investigate the physical conditions in the circumstellar environments, through the estimates of parameters such as gas electron density $n_{\text e}$, electron temperature $T_{\text e}$ and state of ionisation. 

In addition to the estimation of the current mass-loss rate, radio observations of the thermal continuum make it possible to derive fundamental physical parameters such as ionised mass, electron density, and flux of ionising photons, assuming specific conditions on the source geometry and optically thin conditions \citep{Mezger_1967, Moran_1983}.
The ionised mass can be derived from the radio flux density through the relationship between free-free emission and the amount of ionised gas present in the nebula, while the rate of ionising photons provides information on the ionisation capacity of the central source. The emission measure, defined as the integral of $n_{\text e}^2$ along the line of sight, quantifies the integrated electron density and can be derived directly from radio observations under optically thin conditions. However, these parameters require critical assumptions on source geometry (spherical, shell, bipolar) and density distribution, with spherical or quasi-spherical nebulae representing the simplest morphological class typically associated with relatively isotropic mass loss.

The radio spectral index provides diagnostic information on the dominant emission mechanisms. The thermal free-free emission, characteristic of the ionised regions in thermal equilibrium, shows $\alpha\simeq -0.1$ under optically thin conditions and $\alpha\simeq$ 2 in optically thick, while the synchrotron emission, indicative of relativistic electrons, presents $\alpha<$ 0 \citep{Dulk_85}. The shock emission can vary significantly depending on the conditions, with adiabatic shocks showing $\alpha\simeq-0.5$ and radiative shocks producing more complex spectra {\citep{bell_78, Chevalier_82}. In observational reality, radio spectra of CSEs often present a combination of emission mechanisms, resulting in curved spectra for the transition from optically thick to optically thin regimes, multi-component spectra for the superposition of thermal and non-thermal emission, and spatial variations where different mechanisms dominate in different regions of the CSE. A detailed analysis of the observational challenges for detecting these non-thermal components, and the use of spectral index maps to disentangle different emission mechanisms, is provided in Sect.~\ref{sec:csn}. These dia\-gno\-stic capabilities are particularly valuable for bow shock studies, where discriminating between thermal bremsstrahlung and synchrotron emission is crucial for understanding particle acceleration efficiency in stellar wind shocks. For spectral index mapping, it is also key to consider how the observed map may deviate from the intrinsic map due to resolved-out emission and frequency-dependent primary beam sizes; such deviations should be corrected via simulated observations and comparison with other extended objects across the field of view.  

Spatially resolved spectral index mapping could be a powerful diagnostic tool to derive detailed spatial information about physical conditions throughout the CSE. In fact, as the pixel-by-pixel analysis of the SED observed in the mid/far IR has been used to constrain the thermal properties of dust grains and their distribution \citep{Etxaluze_13}, radio spectral index maps constrain the ionisation state, electron temperature, and emission mechanisms of the ionised gas component (e.g. \citealp{Buemi_hr_17,Agliozzo_17}). However, robust constraints on physical parameters require adequate frequency coverage, sufficient angular resolution to separate central components from the surrounding nebulae, and high sensitivity to detect even small variations in the spectral index.

Sect.~\ref{limit_and_potential} highlights how the unprecedented resolution and sensitivity of SKA will allow for detailed mapping of radio morphology in these complex regions.
The integration of multi-wavelength observations spanning radio, infrared, and millimeter enables a complete characterisation of all the components of the CSE.
%
\section{From Progenitors to Supernovae and Remnants: The Legacy of the CSM}
\label{sec:snr}
A legacy of evolved massive stars is the structure and geometry of the circumstellar medium (CSM) that they shape through steady winds, eruptive mass-loss episodes, and binary interactions. When such stars undergo core collapse, the expanding supernova (SN) shock propagates through this structured CSM, affecting the observed spectrophotometric properties of the SN itself. 
The radio continuum, spectral index, and polarization of both SNe and supernova remnants (SNRs) are strongly modulated by the density distribution, clumping, and magnetic fields of the progenitor wind environment. For example, dense shells or clumpy outflows can dominate the early radio light curves of interacting transients such as Type IIn/Ibn SNe, while asymmetric winds and bow shocks leave long-lasting signatures in the morphology and non-thermal emission of young remnants. Delayed interactions, emerging hundreds of days after explosion, trace massive eruptions that occurred millennia before core collapse, whereas rapid post-explosion signatures reveal violent and sporadic mass loss in the years to centuries preceding the SN.

As an example, multi-wavelength monitoring of transitional events such as SN~2014C \citep{Mili2015, Marg2017} has shown how an apparently ordinary Type Ib SN can evolve into an interacting Type IIn once its shock encounters dense, highly asymmetric CSM arranged in equatorial disks or toroidal structures. Such observations point to binary interaction and envelope stripping as key drivers of the pre-SN environment and, ultimately, of the explosion’s observable properties. Another important example is SN~1987A, the best-studied core-collapse SN to date, whose remnant demonstrates with unmatched clarity how a dense and highly inhomogeneous CSM, reflecting the mass loss history of the progenitor, can sculpt the expanding shock and shape its emission across the electromagnetic spectrum, from radio to X-rays. Similarly, SN 1993J (Type IIb) exemplifies the diagnostic potential of radio observations in tracing the interaction between the SN ejecta and the wind from the RSG progenitor, modified by binary mass transfer. Long-term radio monitoring, in fact, has enabled the tracking of the shock front evolution over decades, allowing reconstruction of the progenitor’s mass-loss history, CSM asymmetries, ejecta opacity evolution, and the radial evolution of both the CSM density profile and magnetic fields within the radiating region \citep{Marcaide_09,Peres_2001,Marti-Vidal_2011a,Marti-Vidal_2011b}. Transitional and interacting SNe thus highlight the diagnostic power of radio and X-ray data in probing the CSM within $\sim 0.1$~pc of the progenitor, directly constraining its mass-loss history over decades to millennia before collapse.

We notice that the interaction of shock fronts produced by SN explosions and dense ambient material provides an ideal environment for the production of hadronic $\gamma$-ray emission, and interacting SNe are good candidates for being able to accelerate particles up to PeV energies (e.g., \citealt{2018MNRAS.479.4470M}). Indeed, the inelastic collisions of ultrarelativistic protons accelerated at the shock front with ambient protons produce pions, which decay into $\gamma$-rays and neutrinos (e.g. \citealt{2019ApJ...874...80M}). In this framework, SKA will play a pivotal role in the multi-wavelength and multi-messenger studies of these systems by providing important constraints for future $\gamma$-ray observatories (as CTAO) and neutrino telescopes (as KM3NeT).

In parallel, LBV stars (see Sect.~\ref{sec:LBV}) provide another channel where extreme pre-SN mass loss sculpts dense and asymmetric environments. Their violent, episodic eruptions often produce nested shells or equatorial structures (e.g., \citealt{Umana_g26_12}) that strongly influence the subsequent SN shock propagation. 3D hydrodynamic models of of SNRs from LBV-like progenitors \citep{Usta2021} show that such configurations lead to elongated ejecta, jet-like structures, and chemical stratification that persist for thousands of years. These results highlight how SNR morphology can retain clear fingerprints of LBV progenitors, offering a powerful diagnostic of this evolutionary pathway.

In this context, 3D magneto-hydrodynamic (MHD) simulations of SN shocks interacting with complex CSM (including winds, shells, clumps, and disk-like structures sculpted by the progenitor system) provide a crucial link between progenitor evolution and post-explosion observables. Recent modelling efforts (e.g., \citealt{Orla2020, Orla2021, Orla2022, Orla2025}) have demonstrated how variations in mass-loss history and binary interaction can reproduce the diversity of observed SN/SNR morphologies. Such shock-cloud interactions are observed across different SNR types, including Type Ia SNRs such as Kepler, SN 1006, and Tycho \citep{2016A&A...593A..26M}, demonstrating that ambient inhomogeneities play a crucial role regardless of the progenitor nature. Fully 3D simulations of SN~1987A or SN~2014C, for instance, capture the transition from explosion to strong interaction with inhomogeneous CSM, successfully connecting the observed multi-band emission to the progenitor’s mass-loss history and binary configuration \citep{Orla2020, Orla2024}. In the radio band, significant differences in spectral index gradients and polarization signatures are expected (e.g., \citealt{Orla2019, Petruk2023}). Synthetic radio maps derived from such models suggest that the SKA will have the sensitivity and angular resolution needed to disentangle ejecta–CSM interactions, detect anisotropies and clumping, and trace magnetic field amplification. In this framework, the CSM emerges as the pivotal bridge between massive stars and their explosive aftermath. By combining high-resolution SKA observations, including SKA-VLBI, with state-of-the-art 3D MHD simulations, we will be able to unravel this connection and reconstruct, with unprecedented detail, the life and death of massive stars.

\section{Current Observational Limits and the Potential of SKA}
\label{limit_and_potential}
The current radio facilities, such as Jansky Very Large Array (JVLA), MeerKAT, Australia Telescope Compact Array (ATCA), and ASKAP, 
bring complementary strengths to CSE observations, regardless of some specific limitations affecting each of them. 
For example, although ASKAP and MeerKAT offer excellent sensitivity in the ranges of approximately 700~MHz-1.8 GHz and 0.6-3.5 GHz respectively, they both provide
a relatively narrow frequency window for spectral index measurements, needed to analyse emission mechanisms and  opacity effects. 
Although ATCA offers a wider frequency coverage with observing bands ranging from about 1~GHz to 105~GHz, its sensitivity decreases at lower frequencies and suffers limitations in terms of instantaneous bandwidth and continuous frequency coverage. In addition, the linear arrangement of its antennas requires multi-epoch synthesis observations for adequate UV coverage, making it difficult to obtain reliable images for circumstellar nebulae within reasonable time, especially for structures with complex or asymmetrical morphologies. 
The JVLA antenna configuration instead provides a more uniform coverage of the UV plane, which is ideal for imaging extended structures but is less sensitive compared to the SKA-Mid. As an example, assuming an integration time of 20 minutes in A-configuration, JVLA at L-band (observing frequency 1.5 GHz, available bandwidth unaffected by RFI about 600 MHz wide) achieves an rms of about 20~$\mu$Jy/beam\footnote{Estimates based on JVLA Exposure Calculator Tool (ECT) available at https://obs.vla.nrao.edu/ect/}, that is approximately 8 times higher than expected, under the same observing conditions (i.e. same integration time and bandwidth, and setting Briggs image weighting), for SKA-Mid in AA4 configuration using the Band 2 receiver\footnote{Estimates based on SKAO sensitivity calculator available at https://sensitivity-calculator.skao.int/}. At higher frequencies, SKA-Mid Band 5a reaches 1~$\mu$Jy/beam (central frequency 6.55 GHz, continuum bandwidth 3.9 GHz wide, adopted observing time duration 20 minutes, image weighting "Briggs", with robust parameters fixed 0 and no tapering), still outperforming JVLA by a factor of $\sim\,$4 (estimated using the same time on source with a representative frequency fixed at 6 GHz, that is close to the central frequency of the Band 5a receiver and compatible with a bandwidth 3.9 GHz wide).
In addition, JVLA is geographically limited to the Northern Hemisphere, preventing the observation of many target regions in the southern Galactic Plane.

The effect of these limitations determines several scientific challenges.
Current studies are often limited to relatively small samples, which may not be representative of the entire population, making it difficult to draw an overall picture of the mass-loss mechanisms and their dependence on stellar parameters or environmental conditions.
Evolutionary phases with lower mass-loss rates or wind densities, such as quiescent phases of LBV (crucial to understanding the mechanism behind their eruptions) or some subtypes of WR stars are 
difficult to observe with current instruments, and this limits our ability to track the evolution of mass loss during the stellar lifetime. In addition, the limited sensitivity also makes it extremely
 difficult to detect small variations in the radio flux or spectral index that could indicate small changes in wind density, its structure (clumping), or 
 the presence of weak shocks.

Massive stars and their mass loss in outer galaxies are extremely difficult to study. This limits our understanding of how mass loss and stellar 
evolution vary with metallicity, a crucial parameter influencing stellar winds. For example, observations of WR stars in the Small and Large 
Magellanic Clouds have shown differences in their spectra and mass-loss rates compared to their galactic counterparts \citep{Hainich_14}. These studies at radio frequencies are limited by the difficulty in obtaining high-quality data at these distances.

On the other hand, poor frequency coverage limits the precision of spectral index determination, which requires measurements over a wide range of frequencies. As discussed, this is a crucial parameter for many aspects, such as distinguishing the mechanisms of emission (thermal/non-thermal), analysing opacity effects and parameter variability.

In addition to sensitivity and frequency coverage, angular resolution is necessary to distinguish blended sources, disentangle compact central winds from nebular components, but also characterise on a fine scale the circumstellar structure. The existing instruments face difficulties in providing both high angular resolution and sensitivity to the large-scale structure needed for a full star and circumstellar environment characterisation. The achievement of these objectives is often very time-consuming in terms of overall observation request (e.g. uv-coverage, different array configuration, frequency switching, etc.).
Exploiting the full potential of the SKA for studies of evolved massive stars will require a synergistic approach, combining wide-area surveys to uncover new populations and rare
objects with targeted follow-up observations to characterise their winds and CSEs, in detail. SKA-Mid will provide unprecedented angular resolution and surface-brightness sensitivity, while SKA-Low will probe the low-frequency non-thermal emission associated with shocks. Early Science observations and, subsequently, SKA Phase 2 will progressively extend these capabilities, enabling systematic studies across the full range of evolved massive star populations.

\subsection{SKA Potential: CSE Simulated Observations}
\label{sec:csn}
SKA is designed to have a sensitivity and angular resolution that will allow, together with the wide frequency coverage, to tackle all the discussed issues. Its hybrid configuration, with antennas distributed in both compact and extended configurations, optimizes UV plane coverage for a wide range of angular scales, from point sources to extended structures.

\begin{figure}
    \centering
\includegraphics[width=0.49\columnwidth]{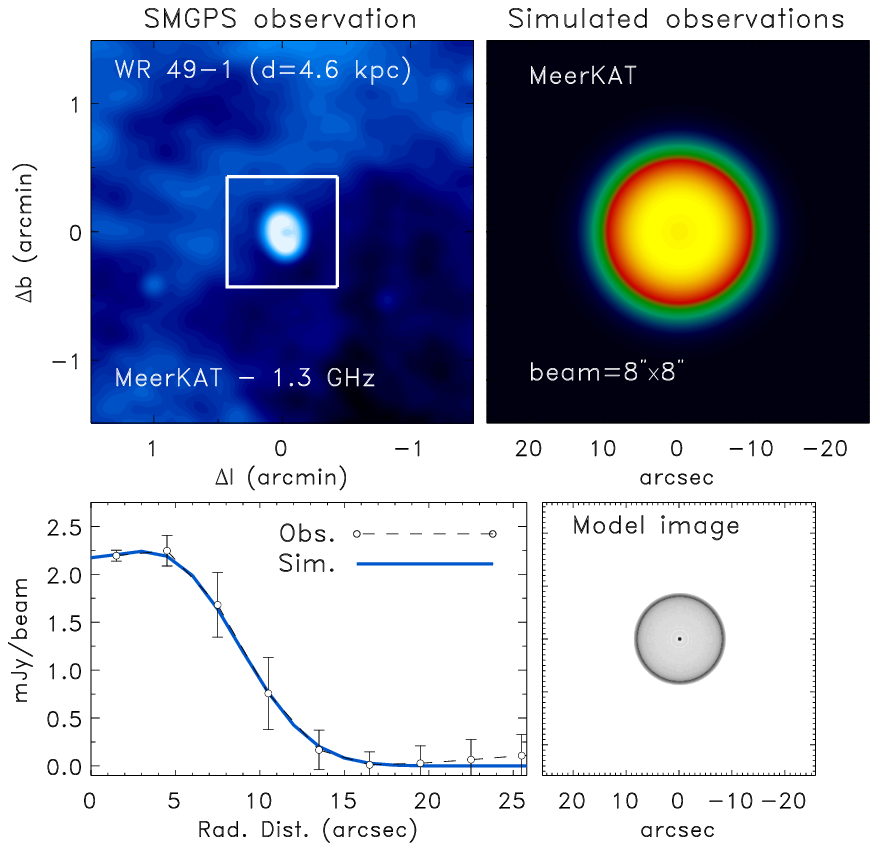}
\hfill
\includegraphics[width=0.49\columnwidth]{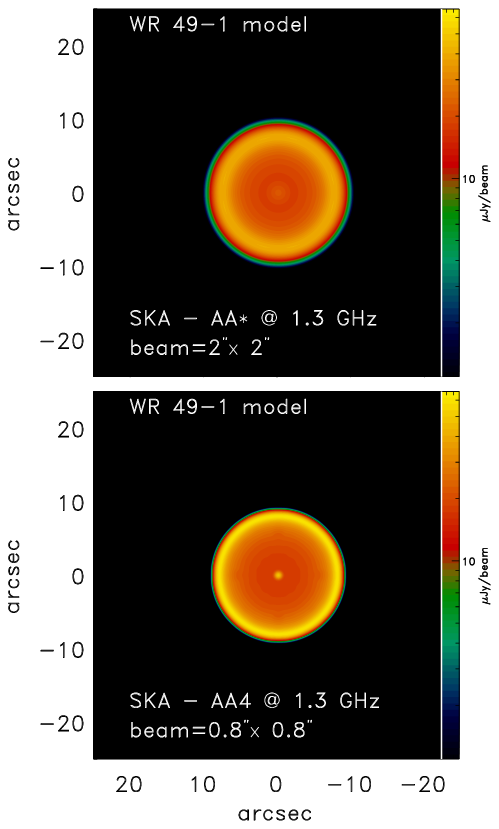}   
\caption{ WR~49-1 nebula as observed in SMGPS, compared with MeerKAT and SKA-Mid synthetic images. {\it Left column, top row:} SMGPS observation at 1.3 GHz (left) and  best-fit model convolved with the corresponding MeerKAT beam (8$^{\prime \prime}$, right). {\it Left column, bottom row:} Radial brightness profile with best-fitting spherical shell model overlaid (left), and the unconvolved model image including an added compact central source (right). {\it Right column:} Same model, convolved with the expected SKA-Mid beam in AA*  (2$^{\prime \prime}$, top) and AA4 (0.8$^{\prime \prime}$, bottom). See Sect.~\ref{sec:csn} for details on the simulation procedures and synthesized beam parameters.}
 \label{fig:wr49-1_sim}
\end{figure}

To obtain a predictive evaluation of the expected performances, we produced synthetic observations based on simplified 3D thermal shell models fitted to real data 
from the SARAO MeerKAT
1.3\,GHz Galactic Plane Survey (SMGPS; \citealt{Goedhart_24}) and then convolved with the expected synthesised beam of the considered arrays.
The objective is to simulate the expected SKA-Mid response to well-characterised circumstellar nebulae, and to compare 
the detectability and morphological fidelity of extended structures and embedded components under different observational conditions and array configurations.
The modeling procedure, originally described in \cite{Umana_2008} and applied to SMGPS observations of WR nebulae in \cite{Buemi_mgps_inprep}, calculates free-free radio emission from a spherically symmetric shell by integrating the radiative transfer equation along lines of sight through a cubic volume. The electron density is assumed to follow a truncated power-law radial distribution. For each source, the model parameters (inner and outer radii, electron density) were constrained by fitting the observed radial brightness profile from the SMGPS maps. The resulting model image represents a first-order approximation of the source structure and provides a physically motivated baseline for instrument comparison. The model brightness distribution was then convolved separately with the MeerKAT beam (8$^{\prime \prime}$) and the expected SKA-Mid beam in both AA* (2$^{\prime \prime}$) and AA4 (0.8$^{\prime \prime}$) configurations at Band 2. 
These synthetic images are intended to provide an illustrative comparison of angular resolution
and surface-brightness sensitivity achievable with current and future instrumentation, rather than to reproduce full interferometric imaging or deconvolution effects.

As representative scientific case, we considered the WR stars WR~49-1 and WR~75ab and their associated extended radio nebulae, both well detected in the SMGPS  \citep{Buemi_mgps_inprep}. These sources are representative of the WR population that will be systematically surveyed by the proposed 10–15~GHz GP survey 
\citep{Traficante01.2026.SKA}
In the current MeerKAT map of WR~49-1, the emission is dominated by the diffuse structure, and it is not possible to assess whether a central source is present or not. So we add to the model a radio source of 30~mJy/pixel in flux (e.g. representing a stellar wind), located in the modeled map of the shell in the central pixel (theoretically adopted pixel size 0.15 arcsec) of the matrix, and then convolved the source model with both MeerKAT and SKA-Mid beams. 

As shown in Fig.~\ref{fig:wr49-1_sim}, the AA* configuration already provides a significant improvement over MeerKAT, partially resolving the shell structure. 
The full AA4 configuration further enhances the angular resolution, enabling robust detection
of the central star’s radio emission, which is completely blended or lost in the MeerKAT image. This result is critical for identifying the ongoing stellar wind contribution, separating it from older ejecta, and constraining current mass loss. The WR49-1 nebula, as observed in SMGPS, is compared with MeerKAT and SKA-Mid synthetic images, convolved with the corresponding instrumental beams. A compact central source is added to test its detectability with SKA.

 \begin{figure}
    \centering
 \includegraphics[width=0.49\columnwidth]{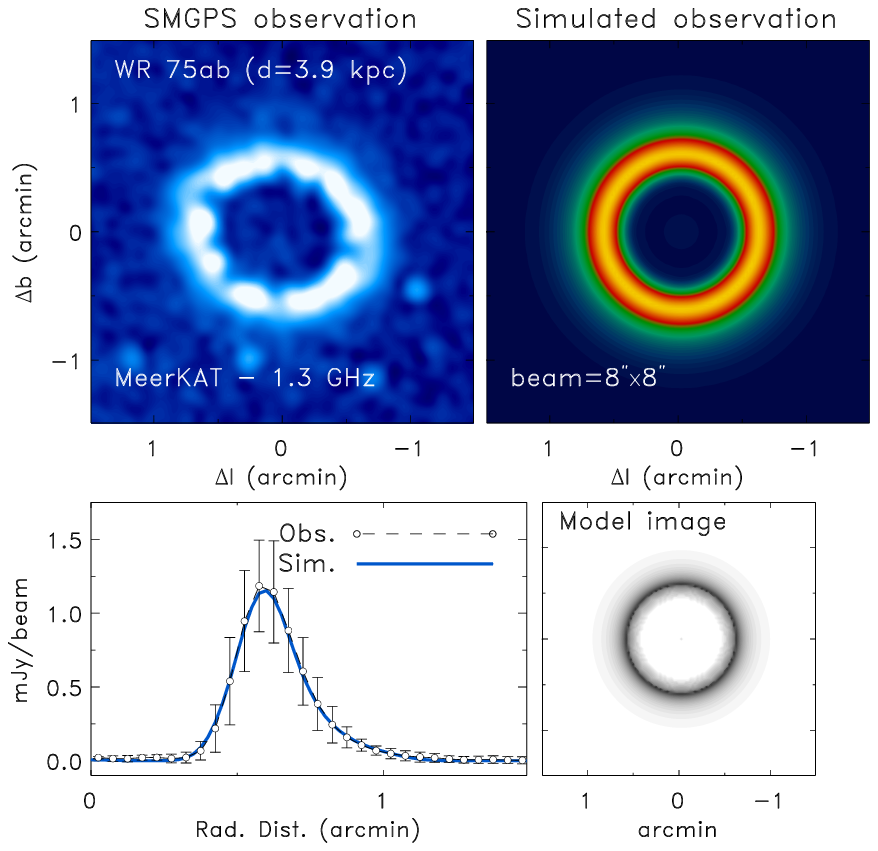}
\hfill
\includegraphics[width=0.49\columnwidth]{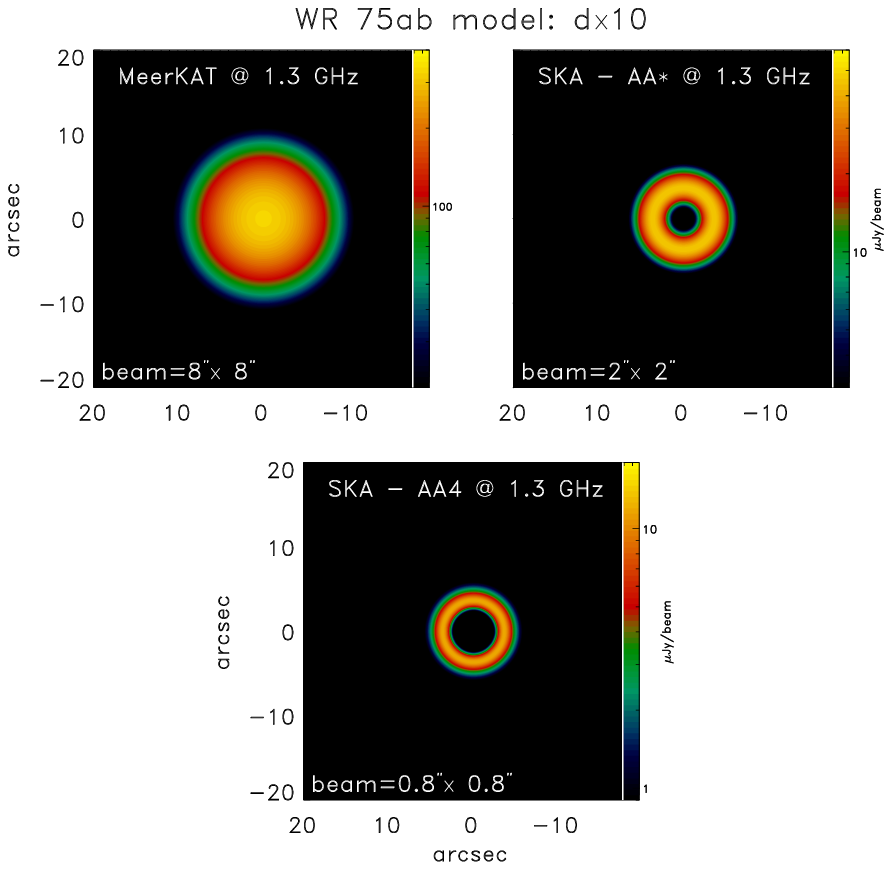}   
 
\caption{ SKA detectability test for the WR 75ab nebula at extragalactic distances. {\it Left column:} SMGPS observation and best-fit model (same procedure as Fig.~\ref{fig:wr49-1_sim}). {\it Right column:} Synthetic images of the same model relocated at $\sim$39 kpc, convolved with MeerKAT beam (unresolved, top-left), SKA-Mid AA* beam (2$^{\prime \prime}$, marginally resolved, top-right), and SKA-Mid AA4 beam (0.8$^{\prime \prime}$, shell morphology fully recovered, bottom).
}
    \label{fig:wr75ab_sim}
\end{figure}

\begin{figure}
    \centering
    \includegraphics[width=0.99\columnwidth]{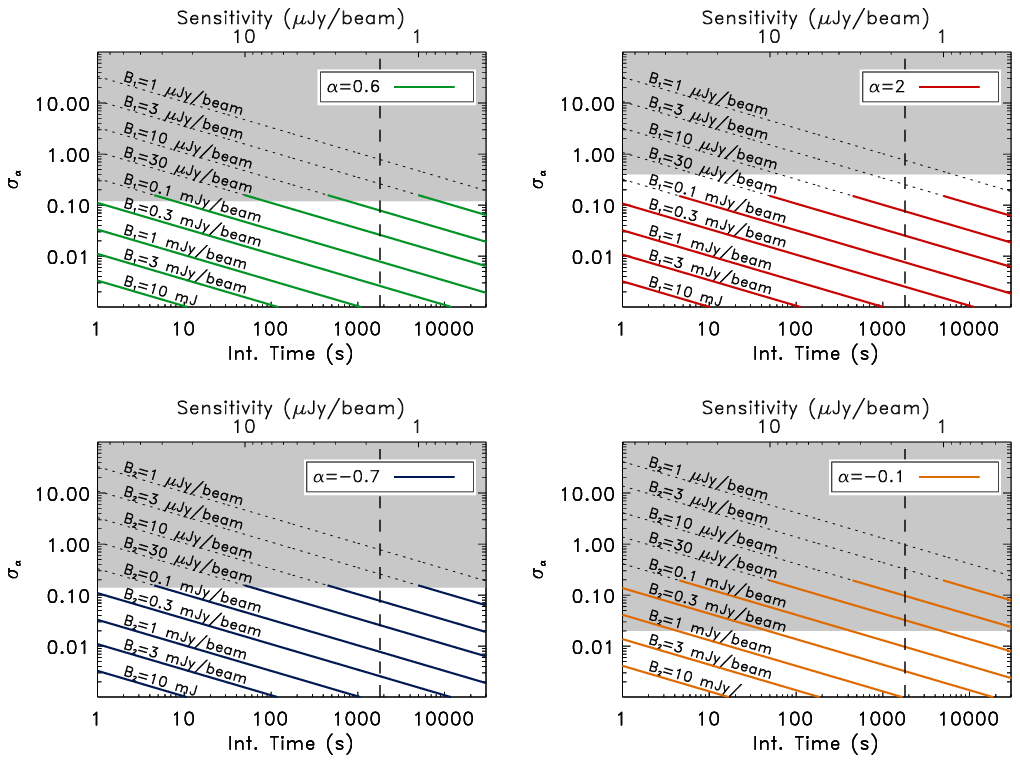}
    \caption{Error on spectral index ($\sigma_{\alpha}$) between SKA-Mid Band~2 (central frequency 1.31~GHz) and Band~5b (central frequency 11.85~GHz) as a function of integration time (sensitivity) for sources of different brightness levels ($B$). 
    Top panels refer to extended radio emission with positive spectral indices ($B_1 < B_2$). Bottom panels refer to cases where the spectral indices of the brightness are negative ($B_2 < B_1$). 
    Coloured thick solid lines highlight detections above the $3 \sigma$ threshold. Grey shaded areas refer to low-reliability estimates of $\alpha$; uncertainty on $\alpha$ higher than 20\%. The dashed vertical line indicates the 30-minute integration time, typical of blind surveys.
}
    \label{fig:digram_diagn_spinde}
\end{figure}

The nebula around WR~75ab offers a second useful illustrative case. As for WR~49-1, the source radial brightness profile has been fitted with a thermal emitting shell model. Then it has been relocated ten times farther away, at $\sim$39 kpc, approximately corresponding to the distance of the Magellanic Clouds. Already with the AA* configuration, the shell becomes marginally resolved, while SKA-Mid AA4  will have enough resolution to reconstruct its morphology (Fig.~\ref{fig:wr75ab_sim}), allowing studies of CSM structures in Local Group galaxies, and thus opening the possibility for population studies beyond the Milky Way.

These simulations demonstrate that even during the Early Science phase (AA*), SKA-Mid will deliver transformative improvements over current facilities. The progression to AA4 and eventually Phase 2 will progressively extend such analyses to intrinsically fainter systems, larger distances, and finer spatial scales within circumstellar nebulae.

Of course, as the spatial resolution increases, the observed  flux is distributed over a larger number of beams, and thus the pixel-to-pixel signal-to-noise ratio decreases. This can limit the capability to derive a reliable local spectral index and investigate its variations across the CSE surrounding the star. In Fig.~\ref{fig:digram_diagn_spinde} is reported the predicted uncertainty on the spectral index typical of different physical emission mechanisms, as a function of the expected sensitivity (and thus the needed integration time) as derived from the sensitivity calculator, assuming different source brightness.
The choice of Band~2 (0.95--1.76 GHz) and Band~5b (4.6--13.8 GHz) provides optimal frequency leverage for spectral index determination while maintaining sensitivity to both compact stellar winds and extended nebular structures.

For a reliable estimate of the spectral index maps of extended radio sources, a necessary condition is to compare radio maps obtained at different frequencies and having similar beam sizes at both bands. 
We therefore used the SKAO Sensitivity Calculator to identify appropriate combinations of input parameters producing nearly matched beams in Band 2 and Band 5b ($\sim$0.8$^{\prime\prime}$).
To perform the estimations, we assumed the full potential of SKA-Mid (AA4 sub-array configuration). For Band~2, we used the Briggs weighting, with a robustness parameter equal to 0, and without tapering. For the Band~5b, we consider as outputs of the SKAO Sensitivity Calculator those obtained by applying a Gaussian taper to the UV data, with a tapering value of $0.473^{\prime\prime}$.
The expected total continuum sensitivities were computed for integration times up to 8 hours.
 The uncertainties of the spectral index, for sources with different pixel brightness, have been calculated using standard error propagation
and for several typical spectral indices.
The plot shows the SKA capability to characterise emission structures on small angular scales, needed for detailed morphological and physical diagnostics of CSE, as well as to obtain reliable $\alpha$ measurements also for faint  compact or unresolved sources, allowing to efficiently perform these studies without the need for concatenating heterogeneous datasets from multiple facilities, with different characteristics and calibrations. This will improve the reliability of spectral index determinations and significantly shorten the data acquisition and analysis process.

\subsection{SKA Potential: Bow Shock Detectability Predictions}
\label{ska:bow_s}

Estimating the future detectability of runaway bow shocks is challenging for a range of regions. From a modeling point of view, many parameters affect the final radio surface brightness: for thermal (free-free) radio emission, the temperature and electron density set the expected emissivity; for non-thermal (synchrotron) radio emission, the shock magnetic field, stellar wind power, particle acceleration efficiency, and slope of the particle energy distribution, also come into play. Furthermore, both emission processes depend on the geometry (in physical and angular units) of the shock, which therefore further depend on the ISM and stellar wind properties. As a final difficulty, the presence of other (extended) radio sources may complicate the detection of an otherwise detectable bow shock. 

Nonetheless, we can use the properties of known infrared bow shocks to make educated estimates of the detection fraction of radio bow shocks. For this purpose, we base our calculations on the E-BOSS survey \citep{2012A&A...538A.108P,2015A&A...578A..45P}, which contains estimates of the shock geometry and ISM density for its sample of $<100$ IR bow shocks. For these systems, we make a number of simplifying assumptions: a $10\%$ acceleration efficiency, a particle energy distribution slope $p=2$, a maximum particle energy of $10^{13}$~eV, and a relatively large magnetic field: e.g., the shock-amplified magnetic field close to the maximum value that allows for a compressible (i.e., shock-forming) ISM, but not exceeding $\sim 25$ $\mu$G. For the expected thermal flux density, we assume an ISM temperature of the order $10^4$ K and a typical stellar velocity of $30$ km/s (note that a lower velocity implies a higher ISM density for a given geometry and therefore a higher thermal emissivity). We finally used a typical beam size of $15$ arc seconds and assumed 1.4 GHz (SKA-mid) and 300 MHz (SKA-low) observing frequencies. 

\begin{figure}
    \centering
    \includegraphics[width=\textwidth]{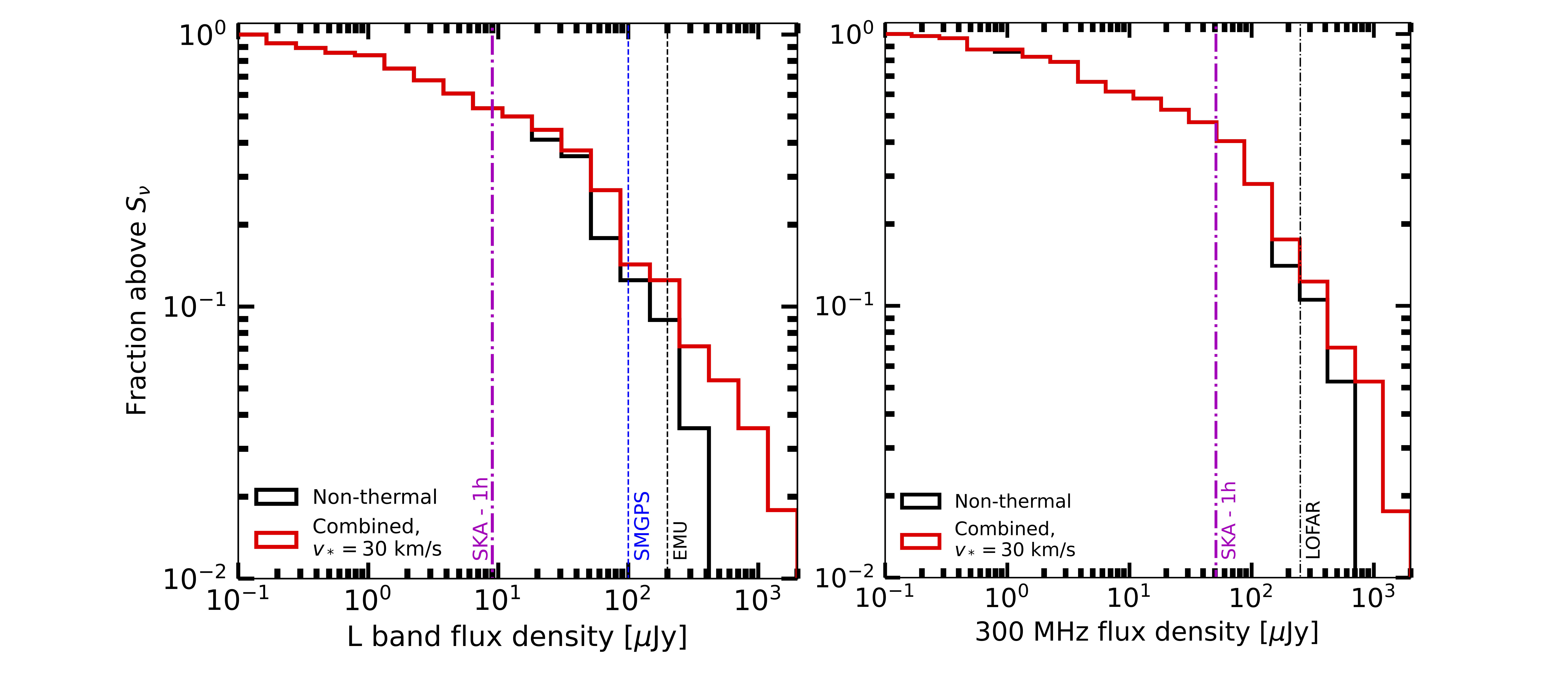}
    \caption{The fraction of bow shocks expected to be detectable with SKA-mid (left) and SKA-low (right). We assume detection thresholds of $\sim 9 / 50$ $\mu$Jy for SKA-mid/low, respectively, versus typical detection thresholds in current surveys of $\sim 100/200/250$ $\mu$Jy for SMGPS/EMU/LOFAR.}
    \label{fig:bowshocks}
\end{figure}

In Figure \ref{fig:bowshocks}, we show the expected fraction of detectable radio bow shocks for a given detection threshold (SKA-mid, left; SKA-low, right). We emphasise that these numbers should be taken as orders of magnitude and that, with the SKA, the measurement of these fractions can be used to test the assumptions made in calculating them. In both bands a significant increase of detectable sources can be expected: at L band (SKA-mid), up to $\sim 50\%$ of the source may have detectable radio emission. Given the currently $\sim 10^3$ IR bow shocks and candidates known \citep{2016ApJS..227...18K,2019MNRAS.488.1141J}, and $\sim 75\%$ with negative declination, that translates to several hundreds of radio bow shocks. With such numbers, the statistical properties of the shocks and, e.g., the acceleration efficiency will become possible. Similar detectability increases are expected for SKA-low, although currently bow shocks are not routinely seen at such low frequencies yet. 

Not only the number of radio bow shocks will increase: as can be seen from the comparison of the non-thermal and combined curves in Figure \ref{fig:bowshocks}, it is evident that especially the number of non-thermal-dominated shocks can be expected to increase. To properly understand the emission’s nature, accurate spectral index maps are required too. From \citet{2024MNRAS.532.2920V}, we see that a signal-to-noise ratio (SNR) of $>5$ per pixel in L and UHF band will yield a spectral index error $\Delta \alpha < 1.2$. To reach $\Delta \alpha < 0.3$, in order to differentiate non-thermal and thermal spectra, SNR$>10$ is necessary. SKA will reach such SNR for bow shocks where currently merely a detection may be expected.

\subsection{Population Studies and Surveys: Classifying Bubbles and Compact Radio Shells}

Beyond the more accurate characterisation of known evolved massive stars, their mass-loss processes and their CSEs, the search for new members of these populations is a crucial and necessary goal. This effort is key for two reasons: first, to deepen our understanding of these final evolutionary stages and their associated phenomenology, and second, to expand the census of high-mass supernova progenitors, a group that is presently exceptionally scarce.

The infrared window has traditionally been the most prolific for the search for new evolved massive stars, with spaceborne observatories like \textit{Spitzer} unveiling a \lq\lq bubbling\rq\rq\, Galactic disk \citep{Churchwell2006} which led to the identification of numerous LBV and WR candidates (e.g., \citealp{Wachter2010, Gvaramadze2010, Mizuno2010}) through the detection of compact and circular shells at 24~$\mu$m.

In recent years, multi-frequency radio spectral index studies in combination with multiwavelength morphological analysis have proven an effective method to distinguish ionised structures surrounding evolved massive stars (such as LBV and WR circumstellar shells) from other extended radio sources, like SNRs, H\textsc{ii} regions or Planetary Nebulae \citep{Ingallinera2014, Ingallinera2016}. In this context, SKA precursors, offering an unprecedented leap forward in terms of sensitivity and imaging fidelity, hold great promise as discovery instruments, potentially providing access to previously unobserved populations of evolved massive stars. These facilities allow for the detection of compact or faint ring-like structures missed by previous surveys, that may trace unrecognized LBVs or WRs. Therefore, SKA precursors may play a role analogous to that of past infrared observatories in completing the census of massive stars.

Indeed, these facilities probe deeper into the low-surface brightness Universe, they are revealing an increasing number of unsuspected structures, many of which display a circular or ring-like morphology akin to that of circumstellar shells.  This potential for discovery was initially demonstrated by the numerous faint \lq\lq low-angular-diameter shells\rq\rq\, first noted by \cite{Heywood2022} in the MeerKAT L-band Galactic Center mosaic. Another key example is the serendipitous discovery of Kyklos (J1802-3353), an $\sim$80 arcsec wide, faint radio ring located 6 deg from the Galactic plane and close in projection to the Galactic center \citep{Bordiu_Kyklos_24}. Kyklos exhibits a thermal spectrum ($\alpha=-0.1\pm0.3$) compatible with the canonical value of an optically thin H\textsc{ii} region, which led to its interpretation as a circumstellar shell around a massive star, likely a Wolf-Rayet.

Similar discoveries can now be made at scale by exploiting large area radio continuum surveys. In this respect, the SMGPS, the deepest L-band survey of the Milky Way to date, offers a rich reservoir of unknown objects, as it is populated by nearly 17000 extended radio sources of which 43\% ($\sim$7000) remain unclassified \citep{Bordiu_25_catalogue}. \cite{Bordiu_25} conducted a systematic search for unclassified compact radio rings, specifically those with an angular radius not exceeding 1 arcmin, combining data from the SMGPS and the Galactic Centre mosaic. Through visual inspection, this work identified 164 radio rings not associated with known Galactic radio emitters. About 19\% of the rings contain a central radio point source, and a multiwavelength analysis revealed that roughly 50$\%$ of the rings display a counterpart in at least one mid-or far-infrared band (8, 24 or 70 $\mu$m). While a large fraction of the sample is thought to be associated with H\textsc{ii} regions, PNe and possibly some extragalactic sources (like galaxies in the Zone of Avoidance or Odd Radio Circle candidates), a subset of the rings show clear infrared extended emission and appear to be positionally coincident with variable stars (including Long Period Variables from \textit{Gaia}). These features make these rings extremely promising mass-loss relic candidates, possibly tracing ionised ejecta around unrecognized LBVs or WR stars.

These works confirm the potential of deep radio observations for detecting new evolved massive stars, especially in highly obscured regions of the Galactic Plane where dust extinction makes the identification in the optical or the infrared unfeasible. Looking ahead, the SKA-Mid, with its superior sensitivity and resolution, will enable more comprehensive searches for new objects while addressing current observational limitations, such as the large uncertainties of spectral indices due to the limited frequency coverage. Specifically, the SKA will allow for spatially resolved spectral analyses to confirm the thermal (free-free) nature of the emission, the detection and characterisation of faint, unresolved central components, and more importantly, population-wide statistical studies of  morphological and emission properties. Such population studies will be greatly enhanced by synergy with the proposed GP survey at 10–15 GHz 
\citep{Traficante01.2026.SKA}which will provide systematic high-frequency detections enabling spectral classification and the identification of non-thermal emitters among the newly discovered candidates. Future sensitivity enhancements towards Phase 2 would  further extend these systematic searches to nearby galaxies, allowing comparative studies of circumstellar shells and  mass-loss signatures across different metallicity environments.

\section{Conclusive Remarks}
\label{synergies}
A comprehensive view of massive stellar evolution and its interplay with the interstellar medium requires a detailed reconstruction of the physical conditions within the CSEs. This includes morphology, kinematics, density and temperature profiles, ionisation state, and chemical composition. This goal cannot be accomplished by radio observations alone, demanding a synergistic, multiwavelength approach. 

In this context, the infrared and millimetre windows are ideal complements. They are sensitive to the dust and the molecular gas components of the CSE, and, together with radio observations tracing the ionised circumstellar gas, enable the independent measurement of the three main ingredients of mass-loss. This results in more accurate mass-loss budgets and robust determinations of the gas-to-dust ratios, while providing a better understanding of the role of dust opacity in shielding stellar UV radiation (key for the survival of molecules) and wind-driving mechanisms.

In fact, combined IR, millimeter, and radio studies have already proven fruitful in the study of evolved massive stars and their nebulae (e.g. \citealp{ Umana_hd16_10,Umna_g79_11, Umana_g26_12,Rizzo_23, Bordiu_20, Buemi_hr_17}). These works have revealed the complex interplay between the different circumstellar dust and gas populations, enabling the reconstruction of mass-loss events chronologies through the analysis of the different circumstellar structures.

As the SKA approaches, frontier facilities such as James Webb Space Telescope (JWST) and Atacama Large Millimetre Array (ALMA) will enable a significant leap forward in understanding the mass-loss processes of evolved massive stars. In the millimetre window, ALMA is already playing a crucial role in mapping the chemical inventory in the envelopes of cool supergiants \citep{Kervella_2018} and revealing molecular ejecta around Yellow Hypergiants, \citep{Wallstrom_2017} and LBV stars \citep{Bordiu_20}, characterised by low $^{12}$CO/$^{13}$CO isotopic ratios indicative of CNO-enriched material. Furthermore, ALMA observations can reveal the distribution of other key chemical tracers. Of special interest are Si-bearing species, like SiO and SiS, which are both the gas-phase building blocks of silicate dust, and the outcomes of its destruction.  Mapping the distribution of these molecules relative to dust structures can pinpoint regions of active dust processing such as in shocks \citep{Bordiu_22}. Looking ahead, the planned Wideband Sensitivity Upgrade\footnote{\url{https://www.eso.org/public/teles-instr/alma/wsu/}} will increase ALMA’s instantaneous bandwidth by up to a factor of 4 while greatly improving sensitivity. This will enable fast and comprehensive molecular line surveys around evolved massive stars, leading to the detection of new species and providing observational insights into the complex circumstellar chemistry of these objects.
Likewise, in the infrared domain, JWST represents a major advance, achieving angular resolution 
($\sim$0.3-1.3$^{\prime\prime}$) comparable to SKA and spectroscopic capabilities (NIRSpec, MIRI-MRS) that will allow PAH features, atomic lines, and molecular tracers (H$_2$) to be spatially mapped across circumstellar nebulae. All together, these facilities will provide new constraints on the geometry, origin and processing of the dust, and on the stratification of the different components, from the hot ionised gas (traced by SKA) to the photodissociation region and neutral/molecular gas (traced by JWST/ALMA) and the dust content.

Furthermore, X-ray observations are particularly valuable for detecting shocked gas in wind-wind interaction regions and bow shocks. The combination of radio continuum and X-ray spectroscopy is useful to study the particle acceleration efficiency and magnetic field amplification.

We also highlight the importance of a dedicated effort to connect SKA studies of evolved massive stars with the interpretation of SNe and SNRs. By combining high-resolution radio imaging of CSM structures with advanced MHD modelling, the SKA will not only constrain the mass-loss history of massive stars (e.g., from LBV-like eruptions to binary-driven envelope stripping) but also provide the essential environmental context needed to decode the physics of stellar explosions and their remnants.

The science goals presented in this chapter can be addressed through a combination of approaches, and are highly complementary to those presented in \cite{Traficante01.2026.SKA}. The wide-area, uniform coverage of that survey will provide an unbiased census of radio-emitting evolved massive stars, first-order spectral index estimates, and serendipitous discovery of new CSE candidates (Section 5.3). Targeted deep observations of known LBV, WR, and RSG systems are needed for a detailed characterisation of individual CSEs, including high-fidelity spectral index mapping, detection of faint embedded compact sources, and optimal uv-coverage matched to individual source structures (Section 5.1). For bow shock studies, both survey data and targeted follow-up of IR-identified candidates will be essential to constrain emission mechanisms and shock properties (Section 5.2).

Finally, by combining high-resolution SKA observations with MHD simulations and complementary multi-wavelength data, we will substantially improve our understanding of the connection between massive star evolution and their explosive death, enabling more robust constraints on how progenitor properties determine the diversity of core-collapse supernovae and remnants.
\\

\bibliographystyle{abbrvnat-maxbibnames4}
\bibliography{chapter}

\begin{thebibliography}{134}
\providecommand{\natexlab}[1]{#1}
\providecommand{\url}[1]{\texttt{#1}}
\expandafter\ifx\csname urlstyle\endcsname\relax
  \providecommand{\doi}[1]{doi: #1}\else
  \providecommand{\doi}{doi: \begingroup \urlstyle{rm}\Url}\fi

\bibitem[{Abbott} et~al.(1981){Abbott}, {Bieging}, and {Churchwell}]{Abbott_81}
D.~C. {Abbott}, J.~H. {Bieging}, and E.~{Churchwell}.
\newblock \emph{\apj}, 250:\penalty0 645--659, Nov. 1981.
\newblock \doi{10.1086/159412}.

\bibitem[{Agliozzo} et~al.(2017){Agliozzo}, {Nikutta}, {Pignata}, {Phillips},
  {Ingallinera}, {Buemi}, {Umana}, {Leto}, {Trigilio}, {Noriega-Crespo},
  {Paladini}, {Bufano}, and {Cavallaro}]{Agliozzo_17}
C.~{Agliozzo} et al.
\newblock \emph{\mnras}, 466\penalty0 (1):\penalty0 213--227, Apr. 2017.
\newblock \doi{10.1093/mnras/stw2986}.

\bibitem[{Agrawal} et~al.(2022){Agrawal}, {Sz{\'e}csi}, {Stevenson},
  {Eldridge}, and {Hurley}]{Agrawal_22}
P.~{Agrawal} et al.
\newblock \emph{\mnras}, 512\penalty0 (4):\penalty0 5717--5725, June 2022.
\newblock \doi{10.1093/mnras/stac930}.

\bibitem[{Aharonian} et~al.(2022){Aharonian}, {Ashkar}, {Backes}, {Barbosa
  Martins}, {Becherini}, {Berge}, {Bi}, {B{\"o}ttcher}, {de Bony de Lavergne},
  {Bradascio}, {Brose}, {Brun}, {Bulik}, {Burger-Scheidlin}, {Cangemi},
  {Caroff}, {Casanova}, {Cerruti}, {Chand}, {Chandra}, {Chen}, {Chibueze},
  {Cristofari}, {Damascene Mbarubucyeye}, {Djannati-Ata{\"\i}}, {Ernenwein},
  {Feijen}, {Fichet de Clairfontaine}, {Fontaine}, {Funk}, {Gabici}, {Gallant},
  {Ghafourizadeh}, {Giavitto}, {Giunti}, {Glawion}, {Glicenstein}, {Goswami},
  {Grondin}, {H{\"a}rer}, {Haupt}, {Hinton}, {H{\"o}rbe}, {Hofmann}, {Holch},
  {Holler}, {Horns}, {Jamrozy}, {Joshi}, {Jung-Richardt}, {Kasai},
  {Katarzy{\'n}ski}, {Katz}, {Kh{\'e}lifi}, {Klu{\'z}niak}, {Komin}, {Kosack},
  {Kostunin}, {Kukec Mezek}, {Lang}, {Le Stum}, {Lemi{\`e}re},
  {Lemoine-Goumard}, {Lenain}, {Leuschner}, {Lohse}, {Luashvili}, {Lypova},
  {Mackey}, {Majumdar}, {Malyshev}, {Marandon}, {Marchegiani}, {Marcowith},
  {Mart{\'\i}-Devesa}, {Marx}, {Maurin}, {Meyer}, {Mitchell}, {Moderski},
  {Mohrmann}, {Montanari}, {Moulin}, {Muller}, {Murach}, {Nakashima}, {de
  Naurois}, {Nayerhoda}, {Niemiec}, {Ohm}, {Olivera-Nieto}, {de Ona Wilhelmi},
  {Ostrowski}, {Panny}, {Panter}, {Parsons}, {Peron}, {Prokhorov},
  {P{\"u}hlhofer}, {Punch}, {Quirrenbach}, {Rauth}, {Reichherzer}, {Reimer},
  {Reimer}, {Renaud}, {Reville}, {Rieger}, {Rowell}, {Rudak}, {Ruiz-Velasco},
  {Sahakian}, {Salzmann}, {Sanchez}, {Santangelo}, {Sasaki}, {Sch{\"u}ssler},
  {Schutte}, {Schwanke}, {Shapopi}, {Specovius}, {Spencer}, {Stawarz},
  {Steenkamp}, {Steinmassl}, {Steppa}, {Sushch}, {Suzuki}, {Takahashi},
  {Tanaka}, {Terrier}, {Thorpe-Morgan}, {Tsirou}, {Tsuji}, {Tuffs}, {Unbehaun},
  {van Eldik}, {van Soelen}, {Vecchi}, {Veh}, {Venter}, {Vink}, {Wagner},
  {White}, {Wierzcholska}, {Wong}, {Zacharias}, {Zargaryan}, {Zdziarski},
  {Zhu}, {Zouari}, {{\.Z}ywucka}, {Blackwell}, {Braiding}, {Burton}, {Cubuk},
  {Filipovi{\'c}}, {Tothill}, and {Wong}]{Aharonian_22}
F.~{Aharonian} et al.
\newblock \emph{\aap}, 666:\penalty0 A124, Oct. 2022.
\newblock \doi{10.1051/0004-6361/202244323}.

\bibitem[{Bell}(1978)]{bell_78}
A.~R. {Bell}.
\newblock \emph{\mnras}, 182:\penalty0 147--156, Jan. 1978.
\newblock \doi{10.1093/mnras/182.2.147}.

\bibitem[{Benaglia} et~al.(2010){Benaglia}, {Romero}, {Mart{\'\i}}, {Peri}, and
  {Araudo}]{2010A&A...517L..10B}
P.~{Benaglia} et al.
\newblock \emph{\aap}, 517:\penalty0 L10, July 2010.
\newblock \doi{10.1051/0004-6361/201015232}.

\bibitem[{Blaauw}(1961)]{Blaauw_61}
A.~{Blaauw}.
\newblock \emph{\bain}, 15:\penalty0 265, May 1961.

\bibitem[{Bonanos} et~al.(2024{\natexlab{a}}){Bonanos}, {Maravelias}, {Yang},
  {Tramper}, {de Wit}, {Zapartas}, {Antoniadis}, {Christodoulou}, and
  {Munoz-Sanchez}]{Bonannos_iaus24}
A.~Z. {Bonanos} et al.
\newblock In J.~{Mackey}, J.~S. {Vink}, and N.~{St-Louis}, editors,
  \emph{Massive Stars Near and Far}, volume 361 of \emph{IAU Symposium}, pages
  447--453, Jan. 2024{\natexlab{a}}.
\newblock \doi{10.1017/S1743921322002782}.

\bibitem[{Bonanos} et~al.(2024{\natexlab{b}}){Bonanos}, {Tramper}, {de Wit},
  {Christodoulou}, {Mu{\~n}oz Sanchez}, {Antoniadis}, {Athanasiou},
  {Maravelias}, {Yang}, and {Zapartas}]{Bonannos_24A&A}
A.~Z. {Bonanos} et al.
\newblock \emph{\aap}, 686:\penalty0 A77, June 2024{\natexlab{b}}.
\newblock \doi{10.1051/0004-6361/202348527}.

\bibitem[{Bordiu} et~al.(2019){Bordiu}, {Rizzo}, and {Ritacco}]{Bordiu_19}
C.~{Bordiu}, J.~R. {Rizzo}, and A.~{Ritacco}.
\newblock \emph{\mnras}, 482\penalty0 (2):\penalty0 1651--1663, Jan. 2019.
\newblock \doi{10.1093/mnras/sty2726}.

\bibitem[{Bordiu} et~al.(2021){Bordiu}, {Bufano}, {Cerrigone}, {Umana},
  {Rizzo}, {Buemi}, {Leto}, {Cavallaro}, {Ingallinera}, {Loru}, {Trigilio}, and
  {Riggi}]{Bordiu_20}
C.~{Bordiu} et al.
\newblock \emph{\mnras}, 500\penalty0 (4):\penalty0 5500--5514, Jan. 2021.
\newblock \doi{10.1093/mnras/staa3606}.

\bibitem[{Bordiu} et~al.(2022){Bordiu}, {Rizzo}, {Bufano}, {Quintana-Lacaci},
  {Buemi}, {Leto}, {Cavallaro}, {Cerrigone}, {Ingallinera}, {Loru}, {Riggi},
  {Trigilio}, {Umana}, and {Sciacca}]{Bordiu_22}
C.~{Bordiu} et al.
\newblock \emph{\apjl}, 939\penalty0 (2):\penalty0 L30, Nov. 2022.
\newblock \doi{10.3847/2041-8213/ac9b10}.

\bibitem[{Bordiu} et~al.(2024){Bordiu}, {Filipovi{\'c}}, {Umana}, {Cotton},
  {Buemi}, {Bufano}, {Camilo}, {Cavallaro}, {Cerrigone}, {Dai}, {Hopkins},
  {Ingallinera}, {Jarrett}, {Koribalski}, {Lazarevi{\'c}}, {Leto}, {Loru},
  {Lundqvist}, {Mackey}, {Norris}, {Payne}, {Rowell}, {Riggi}, {Rizzo},
  {Ruggeri}, {Shabala}, {Smeaton}, {Trigilio}, and
  {Velovi{\'c}}]{Bordiu_Kyklos_24}
C.~{Bordiu} et al.
\newblock \emph{\aap}, 690:\penalty0 A53, Oct. 2024.
\newblock \doi{10.1051/0004-6361/202450766}.

\bibitem[{Bordiu} et~al.(2025{\natexlab{a}}){Bordiu}, {Bufano}, {Umana},
  {Rizzo}, {Spingola}, {Trigilio}, {Loru}, {Filipovic}, {Buemi}, {Cavallaro},
  {Cerrigone}, {Leto}, {Ingallinera}, {Riggi}, {Ruggeri}, {Smeaton}, and
  {Woudt}]{Bordiu_25}
C.~{Bordiu} et al.
\newblock \emph{\mnras}, 543\penalty0 (4):\penalty0 3708--3730, Nov.
  2025{\natexlab{a}}.
\newblock \doi{10.1093/mnras/staf1667}.

\bibitem[{Bordiu} et~al.(2025{\natexlab{b}}){Bordiu}, {Riggi}, {Bufano},
  {Cavallaro}, {Cecconello}, {Camilo}, {Umana}, {Cotton}, {Thompson},
  {Bietenholz}, {Goedhart}, {Anderson}, {Buemi}, {Chibueze}, {Ingallinera},
  {Leto}, {Loru}, {Mutale}, {Rigby}, {Trigilio}, and
  {Williams}]{Bordiu_25_catalogue}
C.~{Bordiu} et al.
\newblock \emph{\aap}, 695:\penalty0 A144, Mar. 2025{\natexlab{b}}.
\newblock \doi{10.1051/0004-6361/202450356}.

\bibitem[Buemi et~al.(2026{\natexlab{a}})Buemi, author2, author3, author4, and
  author5]{Buemi_inprep}
C.~Buemi et al.
\newblock High-resolution radio imaging of circumstellar nebulae around evolved
  massive stars.
\newblock in preparation, 2026{\natexlab{a}}.

\bibitem[Buemi et~al.(2026{\natexlab{b}})Buemi, author2, author3, author4, and
  author5]{Buemi_mgps_inprep}
C.~Buemi et al.
\newblock Meerkat-gps view of wolf-rayet stars and their nebulae.
\newblock in preparation, 2026{\natexlab{b}}.

\bibitem[{Buemi} et~al.(2010){Buemi}, {Umana}, {Trigilio}, {Leto}, and
  {Hora}]{Buemi_iras_10}
C.~S. {Buemi} et al.
\newblock \emph{\apj}, 721\penalty0 (2):\penalty0 1404--1411, Oct. 2010.
\newblock \doi{10.1088/0004-637X/721/2/1404}.

\bibitem[{Buemi} et~al.(2017){Buemi}, {Trigilio}, {Leto}, {Umana},
  {Ingallinera}, {Cavallaro}, {Cerrigone}, {Agliozzo}, {Bufano}, {Riggi},
  {Molinari}, and {Schillir{\`o}}]{Buemi_hr_17}
C.~S. {Buemi} et al.
\newblock \emph{\mnras}, 465\penalty0 (4):\penalty0 4147--4158, Mar. 2017.
\newblock \doi{10.1093/mnras/stw3074}.

\bibitem[{Burgemeister} et~al.(2013){Burgemeister}, {Gvaramadze},
  {Stringfellow}, {Kniazev}, {Todt}, and {Hamann}]{Burgemeister_13}
S.~{Burgemeister} et al.
\newblock \emph{\mnras}, 429\penalty0 (4):\penalty0 3305--3315, Mar. 2013.
\newblock \doi{10.1093/mnras/sts588}.

\bibitem[{Cano-Gonz{\'a}lez} et~al.(2024){Cano-Gonz{\'a}lez}, {Sch{\"o}del},
  {Alberdi}, {Mold{\'o}n}, {P{\'e}rez-Torres}, {Najarro}, and
  {Gallego-Calvente}]{Cano-Gonzalez_24}
M.~{Cano-Gonz{\'a}lez} et al.
\newblock \emph{\aap}, 692:\penalty0 A23, Dec. 2024.
\newblock \doi{10.1051/0004-6361/202451771}.

\bibitem[{Cano-Gonz{\'a}lez} et~al.(2025){Cano-Gonz{\'a}lez}, {Sch{\"o}del},
  {Alberdi}, {Mold{\'o}n}, {P{\'e}rez-Torres}, {Najarro}, and
  {Gallego-Calvente}]{Cano-Gonzalez_25}
M.~{Cano-Gonz{\'a}lez} et al.
\newblock \emph{\aap}, 700:\penalty0 A246, Aug. 2025.
\newblock \doi{10.1051/0004-6361/202554533}.

\bibitem[{Cant{\'o}} et~al.(2000){Cant{\'o}}, {Raga}, and
  {Rodr{\'\i}guez}]{Canto_00}
J.~{Cant{\'o}}, A.~C. {Raga}, and L.~F. {Rodr{\'\i}guez}.
\newblock \emph{\apj}, 536\penalty0 (2):\penalty0 896--901, June 2000.
\newblock \doi{10.1086/308983}.

\bibitem[{Cappa} et~al.(2004){Cappa}, {Goss}, and {van der Hucht}]{Cappa2004}
C.~{Cappa}, W.~M. {Goss}, and K.~A. {van der Hucht}.
\newblock \emph{\aj}, 127\penalty0 (5):\penalty0 2885--2897, May 2004.
\newblock \doi{10.1086/383286}.

\bibitem[{Carretero-Castrillo} et~al.(2023){Carretero-Castrillo}, {Rib{\'o}},
  and {Paredes}]{2023A&A...679A.109C}
M.~{Carretero-Castrillo}, M.~{Rib{\'o}}, and J.~M. {Paredes}.
\newblock \emph{\aap}, 679:\penalty0 A109, Nov. 2023.
\newblock \doi{10.1051/0004-6361/202346613}.

\bibitem[{Chandra}(2018)]{Chandra_18}
P.~{Chandra}.
\newblock \emph{\ssr}, 214\penalty0 (1):\penalty0 27, Feb. 2018.
\newblock \doi{10.1007/s11214-017-0461-6}.

\bibitem[{Chen{\'e}} et~al.(2020){Chen{\'e}}, {St-Louis}, {Moffat}, and
  {Gayley}]{Chene_20}
A.-N. {Chen{\'e}}, N.~{St-Louis}, A.~F.~J. {Moffat}, and K.~G. {Gayley}.
\newblock \emph{\apj}, 903\penalty0 (2):\penalty0 113, Nov. 2020.
\newblock \doi{10.3847/1538-4357/abba24}.

\bibitem[{Chevalier}(1982)]{Chevalier_82}
R.~A. {Chevalier}.
\newblock \emph{\apj}, 258:\penalty0 790--797, July 1982.
\newblock \doi{10.1086/160126}.

\bibitem[{Chisholm} et~al.(2018){Chisholm}, {Tremonti}, and
  {Leitherer}]{Chisholm_18}
J.~{Chisholm}, C.~{Tremonti}, and C.~{Leitherer}.
\newblock \emph{\mnras}, 481\penalty0 (2):\penalty0 1690--1706, Dec. 2018.
\newblock \doi{10.1093/mnras/sty2380}.

\bibitem[{Churchwell} et~al.(2006){Churchwell}, {Povich}, {Allen}, {Taylor},
  {Meade}, {Babler}, {Indebetouw}, {Watson}, {Whitney}, {Wolfire}, {Bania},
  {Benjamin}, {Clemens}, {Cohen}, {Cyganowski}, {Jackson}, {Kobulnicky},
  {Mathis}, {Mercer}, {Stolovy}, {Uzpen}, {Watson}, and
  {Wolff}]{Churchwell2006}
E.~{Churchwell} et al.
\newblock \emph{\apj}, 649\penalty0 (2):\penalty0 759--778, Oct. 2006.
\newblock \doi{10.1086/507015}.

\bibitem[{Cohen} et~al.(2005){Cohen}, {Parker}, and {Green}]{Cohen_05}
M.~{Cohen}, Q.~A. {Parker}, and A.~J. {Green}.
\newblock \emph{\mnras}, 360\penalty0 (4):\penalty0 1439--1447, July 2005.
\newblock \doi{10.1111/j.1365-2966.2005.09137.x}.

\bibitem[{Crowther}(2007)]{Crowther_07}
P.~A. {Crowther}.
\newblock \emph{\araa}, 45\penalty0 (1):\penalty0 177--219, Sept. 2007.
\newblock \doi{10.1146/annurev.astro.45.051806.110615}.

\bibitem[{Daley-Yates} et~al.(2016){Daley-Yates}, {Stevens}, and
  {Crossland}]{Daley2016}
S.~{Daley-Yates}, I.~R. {Stevens}, and T.~D. {Crossland}.
\newblock \emph{\mnras}, 463\penalty0 (3):\penalty0 2735--2745, Dec. 2016.
\newblock \doi{10.1093/mnras/stw2184}.

\bibitem[{Davies} et~al.(2005){Davies}, {Oudmaijer}, and {Vink}]{Davies_05}
B.~{Davies}, R.~D. {Oudmaijer}, and J.~S. {Vink}.
\newblock \emph{\aap}, 439\penalty0 (3):\penalty0 1107--1125, Sept. 2005.
\newblock \doi{10.1051/0004-6361:20052781}.

\bibitem[{Dessart}(2024)]{Dessart_24}
L.~{Dessart}.
\newblock \emph{arXiv e-prints}, art. arXiv:2405.04259, May 2024.
\newblock \doi{10.48550/arXiv.2405.04259}.

\bibitem[{Dougherty} et~al.(2005){Dougherty}, {Beasley}, {Claussen},
  {Zauderer}, and {Bolingbroke}]{Dougherty_2005}
S.~M. {Dougherty} et al.
\newblock \emph{\apj}, 623\penalty0 (1):\penalty0 447--459, Apr. 2005.
\newblock \doi{10.1086/428494}.

\bibitem[{Dougherty} et~al.(2010){Dougherty}, {Clark}, {Negueruela}, {Johnson},
  and {Chapman}]{Dougherty2010AA...511A..58D}
S.~M. {Dougherty} et al.
\newblock \emph{\aap}, 511:\penalty0 A58, Feb. 2010.
\newblock \doi{10.1051/0004-6361/200913505}.

\bibitem[{Drew}(1989)]{Drew_89}
J.~E. {Drew}.
\newblock \emph{\apjs}, 71:\penalty0 267, Oct. 1989.
\newblock \doi{10.1086/191374}.

\bibitem[{Dulk}(1985)]{Dulk_85}
G.~A. {Dulk}.
\newblock \emph{\araa}, 23:\penalty0 169--224, Jan. 1985.
\newblock \doi{10.1146/annurev.aa.23.090185.001125}.

\bibitem[{Eichler} and {Usov}(1993)]{Eichler_Usov1993}
D.~{Eichler} and V.~{Usov}.
\newblock \emph{\apj}, 402:\penalty0 271, Jan. 1993.
\newblock \doi{10.1086/172130}.

\bibitem[{Eldridge} and {Stanway}(2022)]{Eldrige_22}
J.~J. {Eldridge} and E.~R. {Stanway}.
\newblock \emph{\araa}, 60:\penalty0 455--494, Aug. 2022.
\newblock \doi{10.1146/annurev-astro-052920-100646}.

\bibitem[{Etxaluze} et~al.(2013){Etxaluze}, {Goicoechea}, {Cernicharo},
  {Polehampton}, {Noriega-Crespo}, {Molinari}, {Swinyard}, {Wu}, and
  {Bally}]{Etxaluze_13}
M.~{Etxaluze} et al.
\newblock \emph{\aap}, 556:\penalty0 A137, Aug. 2013.
\newblock \doi{10.1051/0004-6361/201321258}.

\bibitem[{Fender} et~al.(2016){Fender}, {Woudt}, {Corbel}, {Coriat}, {Daigne},
  {Falcke}, {Girard}, {Heywood}, {Horesh}, {Horrell}, {Jonker}, {Joseph},
  {Kamble}, {Knigge}, {K{\"o}rding}, {Kotze}, {Kouveliotou}, {Lynch},
  {Maccarone}, {Meintjes}, {Migliari}, {Murphy}, {Nagayama}, {Nelemans},
  {Nicholson}, {O'Brien}, {Oodendaal}, {Oozeer}, {Osborne}, {P{\'e}rez-Torres},
  {Ratcliffe}, {Ribeiro}, {Rol}, {Rushton}, {Scaife}, {Schurch}, {Sivakoff},
  {Staley}, {Steeghs}, {Stewart}, {Swinbank}, {Vergani}, {Warner}, {Wiersema},
  {Armstrong}, {Groot}, {McBride}, {Miller-Jones}, {Mooley}, {Stappers},
  {Wijers}, {Bietenholz}, {Blyth}, {B{\"o}ttcher}, {Buckley}, {Charles},
  {Chomiuk}, {Coppejans}, {de Blok}, {van der Heyden}, {van der Horst}, and
  {van Soelen}]{Fender_16}
R.~{Fender} et al.
\newblock In \emph{MeerKAT Science: On the Pathway to the SKA}, page~13, Jan.
  2016.
\newblock \doi{10.22323/1.277.0013}.

\bibitem[{Flores} and {Hillier}(2021)]{Flores_21}
B.~L. {Flores} and D.~J. {Hillier}.
\newblock \emph{\mnras}, 504\penalty0 (1):\penalty0 311--325, June 2021.
\newblock \doi{10.1093/mnras/stab707}.

\bibitem[{Gallego-Calvente} et~al.(2022){Gallego-Calvente}, {Sch{\"o}del},
  {Alberdi}, {Najarro}, {Yusef-Zadeh}, {Shahzamanian}, and
  {Nogueras-Lara}]{Gallego_22}
A.~T. {Gallego-Calvente} et al.
\newblock \emph{\aap}, 664:\penalty0 A49, Aug. 2022.
\newblock \doi{10.1051/0004-6361/202141895}.

\bibitem[{Garcia-Segura} et~al.(1996){Garcia-Segura}, {Mac Low}, and
  {Langer}]{Garcia-Segura_96}
G.~{Garcia-Segura}, M.-M. {Mac Low}, and N.~{Langer}.
\newblock \emph{\aap}, 305:\penalty0 229, Jan. 1996.

\bibitem[{Goedhart} et~al.(2024){Goedhart}, {Cotton}, {Camilo}, {Thompson},
  {Umana}, {Bietenholz}, {Woudt}, {Anderson}, {Bordiu}, {Buckley}, {Buemi},
  {Bufano}, {Cavallaro}, {Chen}, {Chibueze}, {Egbo}, {Frank}, {Hoare},
  {Ingallinera}, {Irabor}, {Kraan-Korteweg}, {Kurapati}, {Leto}, {Loru},
  {Mutale}, {Obonyo}, {Plavin}, {Rajohnson}, {Rigby}, {Riggi}, {Seidu},
  {Serra}, {Smart}, {Stappers}, {Steyn}, {Surnis}, {Trigilio}, {Williams},
  {Abbott}, {Adam}, {Asad}, {Baloyi}, {Bauermeister}, {Bennet}, {Bester},
  {Botha}, {Brederode}, {Buchner}, {Burger}, {Cheetham}, {Cloete}, {de
  Villiers}, {de Villiers}, {du Toit}, {Esterhuyse}, {Fanaroff}, {Fourie},
  {Gamatham}, {Gatsi}, {Geyer}, {Gouws}, {Gumede}, {Heywood}, {Hokwana},
  {Hoosen}, {Horn}, {Horrell}, {Hugo}, {Isaacson}, {J{\'o}zsa}, {Jonas},
  {Jordaan}, {Joubert}, {Julie}, {Kapp}, {Kriek}, {Kriel}, {Krishnan}, {Kusel},
  {Legodi}, {Lehmensiek}, {Lord}, {Macfarlane}, {Magnus}, {Magozore}, {Main},
  {Malan}, {Manley}, {Marais}, {Maree}, {Martens}, {Maruping}, {McAlpine},
  {Merry}, {Mgodeli}, {Millenaar}, {Mokone}, {Monama}, {New}, {Ngcebetsha},
  {Ngoasheng}, {Nicolson}, {Ockards}, {Oozeer}, {Passmoor}, {Patel},
  {Peens-Hough}, {Perkins}, {Ramaila}, {Ratcliffe}, {Renil}, {Richter},
  {Salie}, {Sambu}, {Schollar}, {Schwardt}, {Schwartz}, {Serylak}, {Siebrits},
  {Sirothia}, {Slabber}, {Smirnov}, {Tiplady}, {van Balla}, {van der Byl}, {Van
  Tonder}, {Venter}, {Venter}, {Welz}, and {Williams}]{Goedhart_24}
S.~{Goedhart} et al.
\newblock \emph{\mnras}, 531\penalty0 (1):\penalty0 649--681, June 2024.
\newblock \doi{10.1093/mnras/stae1166}.

\bibitem[{Groh} et~al.(2013){Groh}, {Meynet}, and {Ekstr{\"o}m}]{Groh_13}
J.~H. {Groh}, G.~{Meynet}, and S.~{Ekstr{\"o}m}.
\newblock \emph{\aap}, 550:\penalty0 L7, Feb. 2013.
\newblock \doi{10.1051/0004-6361/201220741}.

\bibitem[{Gvaramadze} et~al.(2010){Gvaramadze}, {Kniazev}, and
  {Fabrika}]{Gvaramadze2010}
V.~V. {Gvaramadze}, A.~Y. {Kniazev}, and S.~{Fabrika}.
\newblock \emph{\mnras}, 405\penalty0 (2):\penalty0 1047--1060, June 2010.
\newblock \doi{10.1111/j.1365-2966.2010.16496.x}.

\bibitem[{Hainich} et~al.(2014){Hainich}, {R{\"u}hling}, {Todt}, {Oskinova},
  {Liermann}, {Gr{\"a}fener}, {Foellmi}, {Schnurr}, and {Hamann}]{Hainich_14}
R.~{Hainich} et al.
\newblock \emph{\aap}, 565:\penalty0 A27, May 2014.
\newblock \doi{10.1051/0004-6361/201322696}.

\bibitem[Harper et~al.(2001)Harper, Brown, and Lim]{Harper2001}
G.~M. Harper, A.~Brown, and J.~Lim.
\newblock \emph{ApJ}, 551:\penalty0 1073--1098, 2001.
\newblock \doi{10.1086/320215}.

\bibitem[{Heywood} et~al.(2022){Heywood}, {Rammala}, {Camilo}, {Cotton},
  {Yusef-Zadeh}, {Abbott}, {Adam}, {Adams}, {Aldera}, {Asad}, {Bauermeister},
  {Bennett}, {Bester}, {Bode}, {Botha}, {Botha}, {Brederode}, {Buchner},
  {Burger}, {Cheetham}, {de Villiers}, {Dikgale-Mahlakoana}, {du Toit},
  {Esterhuyse}, {Fanaroff}, {February}, {Fourie}, {Frank}, {Gamatham}, {Geyer},
  {Goedhart}, {Gouws}, {Gumede}, {Hlakola}, {Hokwana}, {Hoosen}, {Horrell},
  {Hugo}, {Isaacson}, {J{\'o}zsa}, {Jonas}, {Joubert}, {Julie}, {Kapp},
  {Kenyon}, {Kotz{\'e}}, {Kriek}, {Kriel}, {Krishnan}, {Lehmensiek},
  {Liebenberg}, {Lord}, {Lunsky}, {Madisa}, {Magnus}, {Mahgoub}, {Makhaba},
  {Makhathini}, {Malan}, {Manley}, {Marais}, {Martens}, {Mauch}, {Merry},
  {Millenaar}, {Mnyandu}, {Mokone}, {Monama}, {Mphego}, {New}, {Ngcebetsha},
  {Ngoasheng}, {Ockards}, {Oozeer}, {Otto}, {Passmoor}, {Patel}, {Peens-Hough},
  {Perkins}, {Ramaila}, {Ramanujam}, {Ramudzuli}, {Ratcliffe}, {Robyntjies},
  {Salie}, {Sambu}, {Schollar}, {Schwardt}, {Schwartz}, {Serylak}, {Siebrits},
  {Sirothia}, {Slabber}, {Smirnov}, {Sofeya}, {Taljaard}, {Tasse}, {Tiplady},
  {Toruvanda}, {Twum}, {van Balla}, {van der Byl}, {van der Merwe}, {Van
  Tonder}, {Van Wyk}, {Venter}, {Venter}, {Wallace}, {Welz}, {Williams}, and
  {Xaia}]{Heywood2022}
I.~{Heywood} et al.
\newblock \emph{\apj}, 925\penalty0 (2):\penalty0 165, Feb. 2022.
\newblock \doi{10.3847/1538-4357/ac449a}.

\bibitem[{Huang} et~al.(2023){Huang}, {Jiang}, {Deng}, {Yu}, and
  {Zijlstra}]{Huang2023}
Q.~{Huang} et al.
\newblock \emph{\aj}, 166\penalty0 (1):\penalty0 23, July 2023.
\newblock \doi{10.3847/1538-3881/acd92e}.

\bibitem[{Ignace}(2016)]{Ignace2016}
R.~{Ignace}.
\newblock \emph{\mnras}, 457\penalty0 (4):\penalty0 4123--4134, Apr. 2016.
\newblock \doi{10.1093/mnras/stw216}.

\bibitem[{Ingallinera} et~al.(2014){Ingallinera}, {Trigilio}, {Umana}, {Leto},
  {Noriega-Crespo}, {Flagey}, {Paladini}, {Agliozzo}, and
  {Buemi}]{Ingallinera2014}
A.~{Ingallinera} et al.
\newblock \emph{\mnras}, 437\penalty0 (4):\penalty0 3626--3638, Feb. 2014.
\newblock \doi{10.1093/mnras/stt2157}.

\bibitem[{Ingallinera} et~al.(2016){Ingallinera}, {Trigilio}, {Leto}, {Umana},
  {Buemi}, {Bufano}, {Agliozzo}, {Riggi}, {Flagey}, {Silva}, {Cerrigone}, and
  {Cavallaro}]{Ingallinera2016}
A.~{Ingallinera} et al.
\newblock \emph{\mnras}, 463\penalty0 (1):\penalty0 723--739, Nov. 2016.
\newblock \doi{10.1093/mnras/stw2053}.

\bibitem[{Jayasinghe} et~al.(2019){Jayasinghe}, {Dixon}, {Povich}, {Binder},
  {Velasco}, {Lepore}, {Xu}, {Offner}, {Kobulnicky}, {Anderson}, {Kendrew}, and
  {Simpson}]{2019MNRAS.488.1141J}
T.~{Jayasinghe} et al.
\newblock \emph{\mnras}, 488\penalty0 (1):\penalty0 1141--1165, Sept. 2019.
\newblock \doi{10.1093/mnras/stz1738}.

\bibitem[{Kervella} et~al.(2018){Kervella}, {Decin}, {Richards}, {Harper},
  {McDonald}, {O'Gorman}, {Montarg{\`e}s}, {Homan}, and
  {Ohnaka}]{Kervella_2018}
P.~{Kervella} et al.
\newblock \emph{\aap}, 609:\penalty0 A67, Jan. 2018.
\newblock \doi{10.1051/0004-6361/201731761}.

\bibitem[{Kobulnicky} et~al.(2016){Kobulnicky}, {Chick}, {Schurhammer},
  {Andrews}, {Povich}, {Munari}, {Olivier}, {Sorber}, {Wernke}, {Dale}, and
  {Dixon}]{2016ApJS..227...18K}
H.~A. {Kobulnicky} et al.
\newblock \emph{\apjs}, 227\penalty0 (2):\penalty0 18, Dec. 2016.
\newblock \doi{10.3847/0067-0049/227/2/18}.

\bibitem[{Kotak} and {Vink}(2006)]{Kotak_06}
R.~{Kotak} and J.~S. {Vink}.
\newblock \emph{\aap}, 460\penalty0 (2):\penalty0 L5--L8, Dec. 2006.
\newblock \doi{10.1051/0004-6361:20065800}.

\bibitem[{Lang} et~al.(2005){Lang}, {Johnson}, {Goss}, and
  {Rodr{\'\i}guez}]{Lang_05}
C.~C. {Lang}, K.~E. {Johnson}, W.~M. {Goss}, and L.~F. {Rodr{\'\i}guez}.
\newblock \emph{\aj}, 130\penalty0 (5):\penalty0 2185--2196, Nov. 2005.
\newblock \doi{10.1086/496976}.

\bibitem[{Langer}(2012)]{Langer_12}
N.~{Langer}.
\newblock \emph{\araa}, 50:\penalty0 107--164, Sept. 2012.
\newblock \doi{10.1146/annurev-astro-081811-125534}.

\bibitem[{Leitherer} and {Robert}(1991)]{Leitherer_91}
C.~{Leitherer} and C.~{Robert}.
\newblock \emph{\apj}, 377:\penalty0 629, Aug. 1991.
\newblock \doi{10.1086/170390}.

\bibitem[{L{\'e}pine} and {Moffat}(2008)]{Lepine_08}
S.~{L{\'e}pine} and A.~F.~J. {Moffat}.
\newblock \emph{\aj}, 136\penalty0 (2):\penalty0 548--553, Aug. 2008.
\newblock \doi{10.1088/0004-6256/136/2/548}.

\bibitem[{L{\'e}pine} et~al.(2000){L{\'e}pine}, {Moffat}, {St-Louis},
  {Marchenko}, {Dalton}, {Crowther}, {Smith}, {Willis}, {Antokhin}, and
  {Tovmassian}]{Lepine_2000}
S.~{L{\'e}pine} et al.
\newblock \emph{\aj}, 120\penalty0 (6):\penalty0 3201--3217, Dec. 2000.
\newblock \doi{10.1086/316858}.

\bibitem[{Leung} et~al.(2021){Leung}, {Wu}, and {Fuller}]{leung_21}
S.-C. {Leung}, S.~{Wu}, and J.~{Fuller}.
\newblock \emph{\apj}, 923\penalty0 (1):\penalty0 41, Dec. 2021.
\newblock \doi{10.3847/1538-4357/ac2c63}.

\bibitem[{Lipscy} et~al.(2005){Lipscy}, {Jura}, and {Reid}]{Lipscy_05}
S.~J. {Lipscy}, M.~{Jura}, and M.~J. {Reid}.
\newblock \emph{\apj}, 626\penalty0 (1):\penalty0 439--445, June 2005.
\newblock \doi{10.1086/429900}.

\bibitem[{Maravelias} et~al.(2023){Maravelias}, {de Wit}, {Bonanos}, {Tramper},
  {Munoz-Sanchez}, and {Christodoulou}]{Maravelias_23}
G.~{Maravelias} et al.
\newblock \emph{Galaxies}, 11\penalty0 (3):\penalty0 79, June 2023.
\newblock \doi{10.3390/galaxies11030079}.

\bibitem[{Marcaide} et~al.(2009){Marcaide}, {Mart{\'\i}-Vidal}, {Alberdi},
  {P{\'e}rez-Torres}, {Ros}, {Diamond}, {Guirado}, {Lara}, {Shapiro},
  {Stockdale}, {Weiler}, {Mantovani}, {Preston}, {Schilizzi}, {Sramek},
  {Trigilio}, {van Dyk}, and {Whitney}]{Marcaide_09}
J.~M. {Marcaide} et al.
\newblock \emph{\aap}, 505\penalty0 (3):\penalty0 927--945, Oct. 2009.
\newblock \doi{10.1051/0004-6361/200912133}.

\bibitem[{Marcowith} et~al.(2018){Marcowith}, {Dwarkadas}, {Renaud},
  {Tatischeff}, and {Giacinti}]{2018MNRAS.479.4470M}
A.~{Marcowith} et al.
\newblock \emph{\mnras}, 479\penalty0 (4):\penalty0 4470--4485, Oct. 2018.
\newblock \doi{10.1093/mnras/sty1743}.

\bibitem[{Margutti} et~al.(2017){Margutti}, {Kamble}, {Milisavljevic},
  {Zapartas}, {de Mink}, {Drout}, {Chornock}, {Risaliti}, {Zauderer},
  {Bietenholz}, {Cantiello}, {Chakraborti}, {Chomiuk}, {Fong}, {Grefenstette},
  {Guidorzi}, {Kirshner}, {Parrent}, {Patnaude}, {Soderberg}, {Gehrels}, and
  {Harrison}]{Marg2017}
R.~{Margutti} et al.
\newblock \emph{\apj}, 835\penalty0 (2):\penalty0 140, Feb. 2017.
\newblock \doi{10.3847/1538-4357/835/2/140}.

\bibitem[{Mart{\'\i}-Vidal} et~al.(2011{\natexlab{a}}){Mart{\'\i}-Vidal},
  {Marcaide}, {Alberdi}, {Guirado}, {P{\'e}rez-Torres}, and
  {Ros}]{Marti-Vidal_2011a}
I.~{Mart{\'\i}-Vidal} et al.
\newblock \emph{\aap}, 526:\penalty0 A143, Feb. 2011{\natexlab{a}}.
\newblock \doi{10.1051/0004-6361/201014517}.

\bibitem[{Mart{\'\i}-Vidal} et~al.(2011{\natexlab{b}}){Mart{\'\i}-Vidal},
  {Marcaide}, {Alberdi}, {Guirado}, {P{\'e}rez-Torres}, and
  {Ros}]{Marti-Vidal_2011b}
I.~{Mart{\'\i}-Vidal} et al.
\newblock \emph{\aap}, 526:\penalty0 A142, Feb. 2011{\natexlab{b}}.
\newblock \doi{10.1051/0004-6361/200913831}.

\bibitem[{Massey} et~al.(2000){Massey}, {Waterhouse}, and
  {DeGioia-Eastwood}]{Massey_00}
P.~{Massey}, E.~{Waterhouse}, and K.~{DeGioia-Eastwood}.
\newblock \emph{\aj}, 119\penalty0 (5):\penalty0 2214--2241, May 2000.
\newblock \doi{10.1086/301345}.

\bibitem[{McConnell} et~al.(2020){McConnell}, {Hale}, {Lenc}, {Banfield},
  {Heald}, {Hotan}, {Leung}, {Moss}, {Murphy}, {O'Brien}, {Pritchard}, {Raja},
  {Sadler}, {Stewart}, {Thomson}, {Whiting}, {Allison}, {Amy}, {Anderson},
  {Ball}, {Bannister}, {Bell}, {Bock}, {Bolton}, {Bunton}, {Chippendale},
  {Collier}, {Cooray}, {Cornwell}, {Diamond}, {Edwards}, {Gupta}, {Hayman},
  {Heywood}, {Jackson}, {Koribalski}, {Lee-Waddell}, {McClure-Griffiths}, {Ng},
  {Norris}, {Phillips}, {Reynolds}, {Roxby}, {Schinckel}, {Shields},
  {Tremblay}, {Tzioumis}, {Voronkov}, and {Westmeier}]{McConnell_20}
D.~{McConnell} et al.
\newblock \emph{\pasa}, 37:\penalty0 e048, Nov. 2020.
\newblock \doi{10.1017/pasa.2020.41}.

\bibitem[{Mezger} and {Henderson}(1967)]{Mezger_1967}
P.~G. {Mezger} and A.~P. {Henderson}.
\newblock \emph{\apj}, 147:\penalty0 471, Feb. 1967.
\newblock \doi{10.1086/149030}.

\bibitem[{Miceli} et~al.(2016){Miceli}, {Orlando}, {Pereira}, {Acero},
  {Katsuda}, {Decourchelle}, {Winkler}, {Bonito}, {Reale}, {Peres}, {Li}, and
  {Dubner}]{2016A&A...593A..26M}
M.~{Miceli} et al.
\newblock \emph{\aap}, 593:\penalty0 A26, Aug. 2016.
\newblock \doi{10.1051/0004-6361/201628725}.

\bibitem[{Milisavljevic} et~al.(2015){Milisavljevic}, {Margutti}, {Kamble},
  {Patnaude}, {Raymond}, {Eldridge}, {Fong}, {Bietenholz}, {Challis},
  {Chornock}, {Drout}, {Fransson}, {Fesen}, {Grindlay}, {Kirshner}, {Lunnan},
  {Mackey}, {Miller}, {Parrent}, {Sanders}, {Soderberg}, and
  {Zauderer}]{Mili2015}
D.~{Milisavljevic} et al.
\newblock \emph{\apj}, 815\penalty0 (2):\penalty0 120, Dec. 2015.
\newblock \doi{10.1088/0004-637X/815/2/120}.

\bibitem[{Mizuno} et~al.(2010){Mizuno}, {Kraemer}, {Flagey}, {Billot},
  {Shenoy}, {Paladini}, {Ryan}, {Noriega-Crespo}, and {Carey}]{Mizuno2010}
D.~R. {Mizuno} et al.
\newblock \emph{\aj}, 139\penalty0 (4):\penalty0 1542--1552, Apr. 2010.
\newblock \doi{10.1088/0004-6256/139/4/1542}.

\bibitem[{Moffat} et~al.(1988){Moffat}, {Drissen}, {Lamontagne}, and
  {Robert}]{Moffat_88}
A.~F.~J. {Moffat}, L.~{Drissen}, R.~{Lamontagne}, and C.~{Robert}.
\newblock \emph{\apj}, 334:\penalty0 1038, Nov. 1988.
\newblock \doi{10.1086/166895}.

\bibitem[{Moran}(1983)]{Moran_1983}
J.~M. {Moran}.
\newblock \emph{\rmxaa}, 7:\penalty0 95--107, Aug. 1983.

\bibitem[{Morris} et~al.(2017){Morris}, {Gull}, {Hillier}, {Barlow}, {Royer},
  {Nielsen}, {Black}, and {Swinyard}]{Morris_17}
P.~W. {Morris} et al.
\newblock \emph{\apj}, 842\penalty0 (2):\penalty0 79, June 2017.
\newblock \doi{10.3847/1538-4357/aa71b3}.

\bibitem[{Moutzouri} et~al.(2022){Moutzouri}, {Mackey},
  {Carrasco-Gonz{\'a}lez}, {Gong}, {Brose}, {Zargaryan}, {Toal{\'a}}, {Menten},
  {Gvaramadze}, and {Rugel}]{2022A&A...663A..80M}
M.~{Moutzouri} et al.
\newblock \emph{\aap}, 663:\penalty0 A80, July 2022.
\newblock \doi{10.1051/0004-6361/202243098}.

\bibitem[{Murase} et~al.(2019){Murase}, {Franckowiak}, {Maeda}, {Margutti}, and
  {Beacom}]{2019ApJ...874...80M}
K.~{Murase} et al.
\newblock \emph{\apj}, 874\penalty0 (1):\penalty0 80, Mar. 2019.
\newblock \doi{10.3847/1538-4357/ab0422}.

\bibitem[{Norris} et~al.(2011){Norris}, {Hopkins}, {Afonso}, {Brown}, {Condon},
  {Dunne}, {Feain}, {Hollow}, {Jarvis}, {Johnston-Hollitt}, {Lenc},
  {Middelberg}, {Padovani}, {Prandoni}, {Rudnick}, {Seymour}, {Umana},
  {Andernach}, {Alexander}, {Appleton}, {Bacon}, {Banfield}, {Becker}, {Brown},
  {Ciliegi}, {Jackson}, {Eales}, {Edge}, {Gaensler}, {Giovannini}, {Hales},
  {Hancock}, {Huynh}, {Ibar}, {Ivison}, {Kennicutt}, {Kimball}, {Koekemoer},
  {Koribalski}, {L{\'o}pez-S{\'a}nchez}, {Mao}, {Murphy}, {Messias},
  {Pimbblet}, {Raccanelli}, {Randall}, {Reiprich}, {Roseboom},
  {R{\"o}ttgering}, {Saikia}, {Sharp}, {Slee}, {Smail}, {Thompson}, {Urquhart},
  {Wall}, and {Zhao}]{Norris_11}
R.~P. {Norris} et al.
\newblock \emph{\pasa}, 28\penalty0 (3):\penalty0 215--248, Aug. 2011.
\newblock \doi{10.1071/AS11021}.

\bibitem[{O'Gorman} et~al.(2020){O'Gorman}, {Harper}, {Ohnaka},
  {Feeney-Johansson}, {Wilkeneit-Braun}, {Brown}, {Guinan}, {Lim}, {Richards},
  {Ryde}, and {Vlemmings}]{Gorman20}
E.~{O'Gorman} et al.
\newblock \emph{\aap}, 638:\penalty0 A65, June 2020.
\newblock \doi{10.1051/0004-6361/202037756}.

\bibitem[{Orlando} et~al.(2019){Orlando}, {Miceli}, {Petruk}, {Ono},
  {Nagataki}, {Aloy}, {Mimica}, {Lee}, {Bocchino}, {Peres}, and
  {Guarrasi}]{Orla2019}
S.~{Orlando} et al.
\newblock \emph{\aap}, 622:\penalty0 A73, Feb. 2019.
\newblock \doi{10.1051/0004-6361/201834487}.

\bibitem[{Orlando} et~al.(2020){Orlando}, {Ono}, {Nagataki}, {Miceli}, {Umeda},
  {Ferrand}, {Bocchino}, {Petruk}, {Peres}, {Takahashi}, and
  {Yoshida}]{Orla2020}
S.~{Orlando} et al.
\newblock \emph{\aap}, 636:\penalty0 A22, Apr. 2020.
\newblock \doi{10.1051/0004-6361/201936718}.

\bibitem[{Orlando} et~al.(2021){Orlando}, {Wongwathanarat}, {Janka}, {Miceli},
  {Ono}, {Nagataki}, {Bocchino}, and {Peres}]{Orla2021}
S.~{Orlando} et al.
\newblock \emph{\aap}, 645:\penalty0 A66, Jan. 2021.
\newblock \doi{10.1051/0004-6361/202039335}.

\bibitem[{Orlando} et~al.(2022){Orlando}, {Wongwathanarat}, {Janka}, {Miceli},
  {Nagataki}, {Ono}, {Bocchino}, {Vink}, {Milisavljevic}, {Patnaude}, and
  {Peres}]{Orla2022}
S.~{Orlando} et al.
\newblock \emph{\aap}, 666:\penalty0 A2, Oct. 2022.
\newblock \doi{10.1051/0004-6361/202243258}.

\bibitem[{Orlando} et~al.(2024){Orlando}, {Greco}, {Hirai}, {Matsuoka},
  {Miceli}, {Nagataki}, {Ono}, {Chen}, {Milisavljevic}, {Patnaude}, {Bocchino},
  and {Elias-Rosa}]{Orla2024}
S.~{Orlando} et al.
\newblock \emph{\apj}, 977\penalty0 (1):\penalty0 118, Dec. 2024.
\newblock \doi{10.3847/1538-4357/ad8ac8}.

\bibitem[{Orlando} et~al.(2025){Orlando}, {Janka}, {Wongwathanarat},
  {Bocchino}, {De Looze}, {Milisavljevic}, {Miceli}, {Temim}, {Rho},
  {Nagataki}, {Ono}, {Sapienza}, and {Greco}]{Orla2025}
S.~{Orlando} et al.
\newblock \emph{\aap}, 696:\penalty0 A188, Apr. 2025.
\newblock \doi{10.1051/0004-6361/202553902}.

\bibitem[{Panagia} and {Felli}(1975)]{Panagia_Felli1975}
N.~{Panagia} and M.~{Felli}.
\newblock \emph{\aap}, 39:\penalty0 1--5, Feb. 1975.

\bibitem[{P{\'e}rez-Torres} et~al.(2001){P{\'e}rez-Torres}, {Alberdi}, and
  {Marcaide}]{Peres_2001}
M.~A. {P{\'e}rez-Torres}, A.~{Alberdi}, and J.~M. {Marcaide}.
\newblock \emph{\aap}, 374:\penalty0 997--1002, Aug. 2001.
\newblock \doi{10.1051/0004-6361:20010774}.

\bibitem[{Peri} et~al.(2012){Peri}, {Benaglia}, {Brookes}, {Stevens}, and
  {Isequilla}]{2012A&A...538A.108P}
C.~S. {Peri} et al.
\newblock \emph{\aap}, 538:\penalty0 A108, Feb. 2012.
\newblock \doi{10.1051/0004-6361/201118116}.

\bibitem[{Peri} et~al.(2015){Peri}, {Benaglia}, and
  {Isequilla}]{2015A&A...578A..45P}
C.~S. {Peri}, P.~{Benaglia}, and N.~L. {Isequilla}.
\newblock \emph{\aap}, 578:\penalty0 A45, June 2015.
\newblock \doi{10.1051/0004-6361/201424676}.

\bibitem[{Petruk} et~al.(2023){Petruk}, {Beshley}, {Orlando}, {Bocchino},
  {Miceli}, {Nagataki}, {Ono}, {Loru}, {Pellizzoni}, and {Egron}]{Petruk2023}
O.~{Petruk} et al.
\newblock \emph{\mnras}, 518\penalty0 (4):\penalty0 6377--6389, Feb. 2023.
\newblock \doi{10.1093/mnras/stac3564}.

\bibitem[{Portegies Zwart} et~al.(2010){Portegies Zwart}, {McMillan}, and
  {Gieles}]{Portegie_10}
S.~F. {Portegies Zwart}, S.~L.~W. {McMillan}, and M.~{Gieles}.
\newblock \emph{\araa}, 48:\penalty0 431--493, Sept. 2010.
\newblock \doi{10.1146/annurev-astro-081309-130834}.

\bibitem[{Puls} et~al.(2008){Puls}, {Vink}, and {Najarro}]{Puls_08}
J.~{Puls}, J.~S. {Vink}, and F.~{Najarro}.
\newblock \emph{\aapr}, 16\penalty0 (3-4):\penalty0 209--325, Dec. 2008.
\newblock \doi{10.1007/s00159-008-0015-8}.

\bibitem[{Renzo} et~al.(2017){Renzo}, {Ott}, {Shore}, and {de Mink}]{Renzo_17}
M.~{Renzo}, C.~D. {Ott}, S.~N. {Shore}, and S.~E. {de Mink}.
\newblock \emph{\aap}, 603:\penalty0 A118, July 2017.
\newblock \doi{10.1051/0004-6361/201730698}.

\bibitem[{Rizzo} et~al.(2023){Rizzo}, {Bordiu}, {Buemi}, {Leto}, {Ingallinera},
  {Bufano}, {Umana}, {Cerrigone}, and {Trigilio}]{Rizzo_23}
J.~R. {Rizzo} et al.
\newblock \emph{\aap}, 678:\penalty0 A55, Oct. 2023.
\newblock \doi{10.1051/0004-6361/202346980}.

\bibitem[{Sanchez-Bermudez} et~al.(2019){Sanchez-Bermudez}, {Alberdi},
  {Sch{\"o}del}, {Brandner}, {Galv{\'a}n-Madrid}, {Guirado}, {Herrero-Illana},
  {Hummel}, {Marcaide}, and {P{\'e}rez-Torres}]{Sanchez-Bermudez_2019}
J.~{Sanchez-Bermudez} et al.
\newblock \emph{\aap}, 624:\penalty0 A55, Apr. 2019.
\newblock \doi{10.1051/0004-6361/201834659}.

\bibitem[{Scuderi} et~al.(1998){Scuderi}, {Panagia}, {Stanghellini},
  {Trigilio}, and {Umana}]{Scuderi1998}
S.~{Scuderi} et al.
\newblock \emph{\aap}, 332:\penalty0 251--267, Apr. 1998.

\bibitem[{Smartt}(2015)]{Smartt_15}
S.~J. {Smartt}.
\newblock \emph{\pasa}, 32:\penalty0 e016, Apr. 2015.
\newblock \doi{10.1017/pasa.2015.17}.

\bibitem[{Smartt} et~al.(2009){Smartt}, {Eldridge}, {Crockett}, and
  {Maund}]{Smartt_09}
S.~J. {Smartt}, J.~J. {Eldridge}, R.~M. {Crockett}, and J.~R. {Maund}.
\newblock \emph{\mnras}, 395\penalty0 (3):\penalty0 1409--1437, May 2009.
\newblock \doi{10.1111/j.1365-2966.2009.14506.x}.

\bibitem[{Smith} et~al.(1996){Smith}, {Shara}, and {Moffat}]{Smith_96}
L.~F. {Smith}, M.~M. {Shara}, and A.~F.~J. {Moffat}.
\newblock \emph{\mnras}, 281\penalty0 (1):\penalty0 163--191, July 1996.
\newblock \doi{10.1093/mnras/281.1.163}.

\bibitem[{Smith}(2014)]{Smith_2014}
N.~{Smith}.
\newblock \emph{\araa}, 52:\penalty0 487--528, Aug. 2014.
\newblock \doi{10.1146/annurev-astro-081913-040025}.

\bibitem[{Smith}(2017)]{Smith_17}
N.~{Smith}.
\newblock \emph{Philosophical Transactions of the Royal Society of London
  Series A}, 375\penalty0 (2105):\penalty0 20160268, Sept. 2017.
\newblock \doi{10.1098/rsta.2016.0268}.

\bibitem[{Smith} and {Conti}(2008)]{Smith_Conti}
N.~{Smith} and P.~S. {Conti}.
\newblock \emph{\apj}, 679\penalty0 (2):\penalty0 1467--1477, June 2008.
\newblock \doi{10.1086/586885}.

\bibitem[{Stoop} et~al.(2024{\natexlab{a}}){Stoop}, {de Koter}, {Kaper},
  {Brands}, {Portegies Zwart}, {Sana}, {Stoppa}, {Gieles}, {Mahy}, {Shenar},
  {Guo}, {Nelemans}, and {Rieder}]{2024Natur.634..809S}
M.~{Stoop} et al.
\newblock \emph{\nat}, 634\penalty0 (8035):\penalty0 809--812, Oct.
  2024{\natexlab{a}}.
\newblock \doi{10.1038/s41586-024-08013-8}.

\bibitem[{Stoop} et~al.(2024{\natexlab{b}}){Stoop}, {Derkink}, {Kaper}, {de
  Koter}, {Rogers}, {Ram{\'\i}rez-Tannus}, {Guo}, and
  {Azatyan}]{2024A&A...681A..21S}
M.~{Stoop} et al.
\newblock \emph{\aap}, 681:\penalty0 A21, Jan. 2024{\natexlab{b}}.
\newblock \doi{10.1051/0004-6361/202347383}.

\bibitem[{Toal{\'a}} and {Arthur}(2011)]{Toala2011}
J.~A. {Toal{\'a}} and S.~J. {Arthur}.
\newblock \emph{\apj}, 737\penalty0 (2):\penalty0 100, Aug. 2011.
\newblock \doi{10.1088/0004-637X/737/2/100}.

\bibitem[{Toal{\'a}} et~al.(2015){Toal{\'a}}, {Guerrero}, {Ramos-Larios}, and
  {Guzm{\'a}n}]{Toala2015}
J.~A. {Toal{\'a}}, M.~A. {Guerrero}, G.~{Ramos-Larios}, and V.~{Guzm{\'a}n}.
\newblock \emph{\aap}, 578:\penalty0 A66, June 2015.
\newblock \doi{10.1051/0004-6361/201525706}.

\bibitem[Traficante et~al.(2026)Traficante, author2, author3, author4, and
  author5]{Traficante01.2026.SKA}
A.~Traficante et al.
\newblock In \emph{Advancing Astrophysics with the SKA -- II (AASKAII)}. 2026.
\newblock arXiv search: Report number AASKAII/Traficante01.

\bibitem[{Umana} et~al.(2005){Umana}, {Buemi}, {Trigilio}, and
  {Leto}]{Umana_iras_05}
G.~{Umana}, C.~S. {Buemi}, C.~{Trigilio}, and P.~{Leto}.
\newblock \emph{\aap}, 437\penalty0 (1):\penalty0 L1--L5, July 2005.
\newblock \doi{10.1051/0004-6361:200500126}.

\bibitem[{Umana} et~al.(2008){Umana}, {Trigilio}, {Cerrigone}, {Buemi}, and
  {Leto}]{Umana_2008}
G.~{Umana} et al.
\newblock \emph{\mnras}, 386\penalty0 (3):\penalty0 1404--1410, May 2008.
\newblock \doi{10.1111/j.1365-2966.2008.13044.x}.

\bibitem[{Umana} et~al.(2010){Umana}, {Buemi}, {Trigilio}, {Leto}, and
  {Hora}]{Umana_hd16_10}
G.~{Umana} et al.
\newblock \emph{\apj}, 718\penalty0 (2):\penalty0 1036--1045, Aug. 2010.
\newblock \doi{10.1088/0004-637X/718/2/1036}.

\bibitem[{Umana} et~al.(2011){Umana}, {Buemi}, {Trigilio}, {Leto}, {Agliozzo},
  {Ingallinera}, {Noriega-Crespo}, and {Hora}]{Umna_g79_11}
G.~{Umana} et al.
\newblock \emph{\apjl}, 739\penalty0 (1):\penalty0 L11, Sept. 2011.
\newblock \doi{10.1088/2041-8205/739/1/L11}.

\bibitem[{Umana} et~al.(2012){Umana}, {Ingallinera}, {Trigilio}, {Buemi},
  {Leto}, {Agliozzo}, {Noriega-Crespo}, {Flagey}, {Paladini}, and
  {Molinari}]{Umana_g26_12}
G.~{Umana} et al.
\newblock \emph{\mnras}, 427\penalty0 (4):\penalty0 2975--2984, Dec. 2012.
\newblock \doi{10.1111/j.1365-2966.2012.22018.x}.

\bibitem[{Ustamujic} et~al.(2021){Ustamujic}, {Orlando}, {Miceli}, {Bocchino},
  {Limongi}, {Chieffi}, {Trigilio}, {Umana}, {Bufano}, {Ingallinera}, and
  {Peres}]{Usta2021}
S.~{Ustamujic} et al.
\newblock \emph{\aap}, 654:\penalty0 A167, Oct. 2021.
\newblock \doi{10.1051/0004-6361/202141569}.

\bibitem[{van den Eijnden} et~al.(2022{\natexlab{a}}){van den Eijnden},
  {Heywood}, {Fender}, {Mohamed}, {Sivakoff}, {Saikia}, {Russell}, {Motta},
  {Miller-Jones}, and {Woudt}]{Van_den_Eijnden_vela_22}
J.~{van den Eijnden} et al.
\newblock \emph{\mnras}, 510\penalty0 (1):\penalty0 515--530, Feb.
  2022{\natexlab{a}}.
\newblock \doi{10.1093/mnras/stab3395}.

\bibitem[{van den Eijnden} et~al.(2022{\natexlab{b}}){van den Eijnden},
  {Saikia}, and {Mohamed}]{Van_den_Eijnden_22}
J.~{van den Eijnden}, P.~{Saikia}, and S.~{Mohamed}.
\newblock \emph{\mnras}, 512\penalty0 (4):\penalty0 5374--5389, June
  2022{\natexlab{b}}.
\newblock \doi{10.1093/mnras/stac823}.

\bibitem[{van den Eijnden} et~al.(2024){van den Eijnden}, {Mohamed},
  {Carotenuto}, {Motta}, {Saikia}, and {Williams-Baldwin}]{2024MNRAS.532.2920V}
J.~{van den Eijnden} et al.
\newblock \emph{\mnras}, 532\penalty0 (3):\penalty0 2920--2933, Aug. 2024.
\newblock \doi{10.1093/mnras/stae1622}.

\bibitem[{van den Eijnden} et~al.(2025){van den Eijnden}, {Sidoli}, {Diaz
  Trigo}, {El Mellah}, {Sguera}, {Degenaar}, {F{\"u}rst}, {Grinberg},
  {Kretschmar}, {Mart{\'\i}nez-N{\'u}{\~n}ez}, {Miller-Jones}, {Postnov}, and
  {Russell}]{2025MNRAS.543..862V}
J.~{van den Eijnden} et al.
\newblock \emph{\mnras}, 543\penalty0 (1):\penalty0 862--880, Oct. 2025.
\newblock \doi{10.1093/mnras/staf1525}.

\bibitem[{Van Dyk}(2025)]{VanDyk_25}
S.~D. {Van Dyk}.
\newblock \emph{Galaxies}, 13\penalty0 (2):\penalty0 33, Apr. 2025.
\newblock \doi{10.3390/galaxies13020033}.

\bibitem[{van Marle} et~al.(2015){van Marle}, {Meliani}, and
  {Marcowith}]{van_Marle_15}
A.~J. {van Marle}, Z.~{Meliani}, and A.~{Marcowith}.
\newblock \emph{\aap}, 584:\penalty0 A49, Dec. 2015.
\newblock \doi{10.1051/0004-6361/201425230}.

\bibitem[{Vink}(2012)]{Vink_2012}
J.~S. {Vink}.
\newblock In K.~{Davidson} and R.~M. {Humphreys}, editors, \emph{Eta Carinae
  and the Supernova Impostors}, volume 384 of \emph{Astrophysics and Space
  Science Library}, page 221, Jan. 2012.
\newblock \doi{10.1007/978-1-4614-2275-4_10}.

\bibitem[{Vink}(2022)]{Vink_22}
J.~S. {Vink}.
\newblock \emph{\araa}, 60:\penalty0 203--246, Aug. 2022.
\newblock \doi{10.1146/annurev-astro-052920-094949}.

\bibitem[{Wachter} et~al.(2010){Wachter}, {Mauerhan}, {Van Dyk}, {Hoard},
  {Kafka}, and {Morris}]{Wachter2010}
S.~{Wachter} et al.
\newblock \emph{\aj}, 139\penalty0 (6):\penalty0 2330--2346, June 2010.
\newblock \doi{10.1088/0004-6256/139/6/2330}.

\bibitem[{Wallstr{\"o}m} et~al.(2017){Wallstr{\"o}m}, {Lagadec}, {Muller},
  {Black}, {Cox}, {Galv{\'a}n-Madrid}, {Justtanont}, {Longmore}, {Olofsson},
  {Oudmaijer}, {Quintana-Lacaci}, {Szczerba}, {Vlemmings}, {van Winckel}, and
  {Zijlstra}]{Wallstrom_2017}
S.~H.~J. {Wallstr{\"o}m} et al.
\newblock \emph{\aap}, 597:\penalty0 A99, Jan. 2017.
\newblock \doi{10.1051/0004-6361/201628416}.

\bibitem[{Weis} and {Bomans}(2020)]{Weis_20}
K.~{Weis} and D.~J. {Bomans}.
\newblock \emph{Galaxies}, 8\penalty0 (1):\penalty0 20, Feb. 2020.
\newblock \doi{10.3390/galaxies8010020}.

\bibitem[{White} and {Tuthill}(2026)]{White_Tuthill_26}
R.~M.~T. {White} and P.~{Tuthill}.
\newblock In \emph{Encyclopedia of Astrophysics, Volume 2}, volume~2, pages
  584--603, Jan. 2026.
\newblock \doi{10.1016/B978-0-443-21439-4.00067-5}.

\bibitem[{Wright} and {Barlow}(1975)]{Wright_Barlow1975}
A.~E. {Wright} and M.~J. {Barlow}.
\newblock \emph{\mnras}, 170:\penalty0 41--51, Jan. 1975.
\newblock \doi{10.1093/mnras/170.1.41}.

\bibitem[{Zavala} et~al.(2022){Zavala}, {Toal{\'a}}, {Santamar{\'\i}a},
  {Ramos-Larios}, {Sabin}, {Quino-Mendoza}, {Rubio}, and {Guerrero}]{zavala_22}
S.~{Zavala} et al.
\newblock \emph{\mnras}, 513\penalty0 (3):\penalty0 3317--3325, July 2022.
\newblock \doi{10.1093/mnras/stac1097}.

\end{thebibliography}

\end{document}